\documentclass[journal]{IEEEtranTIE}

\usepackage{amsmath,amsfonts}
\usepackage{graphicx}
\usepackage{float}
\usepackage{booktabs}
\usepackage{array}
\usepackage{multirow}
\usepackage{longtable}
\usepackage{tabularx}
\usepackage{textcomp}
\usepackage{url}
\usepackage{verbatim}
\usepackage{adjustbox}
\usepackage{circuitikz}
\usepackage{stfloats}

\usepackage{algorithm}
\usepackage{algpseudocode}

\usepackage[caption=false,font=normalsize,labelfont=sf,textfont=sf]{subfig}
\usepackage{caption}

\usepackage{cite}

\usepackage{hyperref}

\UseRawInputEncoding

\begin{document}

    % \title{A Literature Survey on Frequency Scanning on IBR Penetrated Power System}
    \title{A Survey on IBR Penetrated Power System Stability Analysis Using Frequency Scanning}
    % \title{A literature survey on IBR Stabality Analysis }
	
	\author{{Shuvangkar Chandra Das, Lokesh Saravana, Le Minh Vu, Manh Bui,  \\
    Tuyen Vu, Jianhua Zhang, Thomas Ortmeyer\\
	\textit{}}
		
		\thanks{Corresponding author: dassc@clarkson.edu}}

	\maketitle
	\vspace{-20pt}

\begin{abstract}

    The rapid rise in inverter-based renewable resources has heightened concerns over subsynchronous resonance and oscillations, thereby challenging grid stability. This paper reviews approaches to identify and mitigate these issues, focusing on frequency scanning methods for stability assessment. It categorizes white-, black-, and gray-box modeling techniques, compares positive-sequence, dq-frame, and alpha-beta domain scanning, and examines perturbation shapes like step, ramp, and chirp. A comparative study highlights their strengths, limitations, and suitability for specific scenarios. By summarizing past events and surveying available tools, this work guides operators and researchers toward more effective, reliable stability analysis methods in grids with high renewable penetration.
\end{abstract}
	

\section{Introduction}

    \IEEEPARstart{T}{he} transition to clean energy is significantly reliant on the adoption of renewable energy sources, which have played crucial role in reducing CO2 emissions. The "Net Zero Emissions by 2050 Scenario" predicts that renewables will contribute to more than a third of the expected reduction in CO2 emissions from 2020 to 2030 \cite{ieaTrackingCleanEnergy2022}. Specifically,the World Energy Commission highlights that each gigawatt-hour (GWh) of wind power can potentially lead to a reduction of up to 600 metric tons of CO2 emissions. \cite{blancoEconomicsWindEnergy2009}. Among various renewable energy options, Inverter-Based Resources (IBR) such as wind power are distinguished for their minimal greenhouse gas emissions, as corroborated by multiple studies \cite{mostafaeipourRenewableEnergyIssues2009, evansAssessmentSustainabilityIndicators2009, blancoEconomicsWindEnergy2009}.

    The ongoing integration of IBRs into the power grid necessitates the gradual retirement of conventional synchronous generators, resulting in a decrease in grid inertia. This phenomenon has sparked concerns regarding the stability of the power system, making it crucial to conduct thorough analyses, particularly in scenarios with high IBR integration.
    
    Sub-synchronous Interaction (SSI), Sub-synchronous Oscillation (SSO), and Sub-synchronous Resonance (SSR) are used interchangeably in most of the current literature. The paper \cite{grossSubSynchronousGridConditions2010} explained  differences among the three terms. SSI is a general term referring to the exchange of energy between two parts of an electrical system with each other at a subsynchronous frequency. SSO is a broad term that defines the results of SSI. SSR is the more specific term when a mechanical mass (synchronous generator) resonates with an effective system impedance. Any device capable of swiftly responding or regulating power and speed fluctuations within the sub-synchronous frequency spectrum could trigger sub-synchronous oscillations \cite{ieeeReadersGuideSubsynchronous1992}.
 
    % \begin{figure}[!htb]
    %     \centering
    %     \includegraphics[width=0.5\textwidth]{Figures/Oscillation_events/oscillation_events_chart.png}
    %     \caption{Number of oscillation events per year (Logarithmic scale)}
    %     \label{fig:oscillation_events_chart}
    % \end{figure}
    
    % \begin{figure}[!htb]
    %     \centering
    %     \includegraphics[width=0.5\textwidth]{Figures/Oscillation_events/oscillation_events_by_country.png}
    %     \caption{Number of oscillation events by country}
    %     \label{fig:oscillation_events_by_country}
    % \end{figure}
    
    % \begin{figure}[!htb]
    %     \centering
    %     \includegraphics[width=0.5\textwidth]{Figures/Oscillation_events/oscillation_events_by_frequency.png}
    %     \caption{Number of oscillation events by frequency}
    %     \label{fig:oscillation_events_by_frequency}
    % \end{figure}

    SSR events are not new to the industry. Butler and Concordia first identified this phenomenon in 1937 \cite{butlerAnalysisSeriesCapacitor1937}. However, it gained significant attention after critical damage was discovered in two shafts of the turbine generator at the Mohave Generating Station in 1970 due to sub-synchronous interaction with the series capacitor. The first paper on sub-synchronous resonance in series compensated transmission lines was presented in 1973 \cite{ballanceSubsynchronousResonanceSeries1973}. In 1992, the IEEE Sub-synchronous Resonance Working Group's report \cite{ieeeReadersGuideSubsynchronous1992} provides comprehensive coverage of the essential concepts, problem definition, analytical tools, testing techniques, and effective countermeasures for mitigating the effects of SSR. 
    % Figure \ref{fig:ssr_ieee} depicted the classification proposed by IEEE. 
    The working group addressed SSO into two categories, such as SSR and DDSSO (Device Dependent Sub-synchronous Oscillation). Further, SSR can be categorized under three different types, such as Induction Generator Effect (IGE), Torsional Interaction (TI), and Torque Amplification (TI). In an actual system, three types of SSR/SSO could exist simultaneously \cite{wangReviewEmergingSSR2017}. Another type of sub-synchronous oscillation that has drawn significant attention due to the increasing integration of IBRs is sub-synchronous control interaction (SSCI). Additionally, some interactions do not have a standardized name. Therefore, Luping Wang et al. proposed a novel classification for subsynchronous resonance/oscillation (SSR/SSO) related to wind power integration through power electronic converters\cite{wangReviewEmergingSSR2017}. Their study explores actual incidents resulting from interactions between wind generators and AC oscillations, illustrating the coexistence and interaction among various SSR/SSO types. Hence, their research advances theoretical understanding in this field and establishes a framework for examining emerging SSR/SSO concerns.

    The stability and reliability of modern power systems are crucial, particularly in the face of complex phenomena such as Subsynchronous Resonance (SSR) and Subsynchronous Oscillations (SSO). These phenomena pose significant risks, potentially leading to catastrophic grid or equipment failures. Addressing these challenges necessitates a comprehensive understanding of their root causes and the development of effective mitigation strategies. A detailed inverter model is essential in this context, as it provides a comprehensive view of possible oscillation frequencies and contributes to a thorough investigation of SSR and SSO. Various methods are available for this purpose, including frequency scans, the Eigenvalue method, electromagnetic transient analysis, and Nyquist stability analysis \cite{elfayoumyComprehensiveApproachSubsynchronous2003}. Among these, frequency scans, categorized as single and multi-frequency scans, play a pivotal role. While single-frequency scans are time-consuming, they are noted for their accuracy and reliability, providing the most accurate results \cite{matsuoOptimizedFrequencyScanning2020}.

    In parallel, the Impedance-Based Stability Analysis (IBSA) method has emerged as a powerful tool for predicting system stability, particularly in large-scale, non-linear grid subsystems that include Inverter-Based Resources (IBRs), High Voltage Direct Current (HVDC) lines, photovoltaic systems, and wind power systems \cite{liaoGeneralRulesUsing2018}. The IBSA method is advantageous because it does not require prior knowledge of the internal parameters of different subsystems, making it a versatile and accessible option for stability assessment. Originally developed for designing the input filters of dc-dc converters, IBSA has found extensive application in power systems stability assessment, offering a solution to the challenges associated with Subsynchronous Stability Assessment (SSBA) \cite{trevisanAnalyticallyValidatedSSCI2021}. The method particularly excels when combined with frequency scanning techniques in Electromagnetic Transient (EMT) software packages or measurements, bypassing the need for complex analytical equation development. Various screening techniques of the grid and Grid-Connected Inverters (GCIs) have been proposed and applied in studies performed by utilities, further validating this approach \cite{sunImpedanceBasedStabilityCriterion2011, badrzadehGeneralMethodologyAnalysis2013, karaagacSafeOperationDFIGBased2018, buchhagenBorWin1FirstExperiences2015a, chengSubsynchronousResonanceAssessment2019, saadResonancesHarmonicsHVDCMMC2017, karnikEvaluationCriticalImpact2017}.

    To apply small signal analysis to the grid subsystem, it must be linearized around a given operating point. The time domain EMT simulation is utilized for this purpose, enabling frequency scanning to identify the impedance model of non-linear IBR resources. Frequency scanning serves as a cost-effective and efficient initial step for SSO analysis, identifying conditions that may lead to induction generator effects, torsional interactions, and torque amplification problems \cite{ieeeReadersGuideSubsynchronous1992}. 
    % While it provides a preliminary screening of potential issues, further detailed studies using other methods are necessary to verify the problem's severity and to determine if these conditions produce damaging levels of torques on the shafts.

    % Frequency scanning being a prevalent tool to analyze the power system stabality, different researchesrs applied this method in different domain such as positive sequences, dq-frame and alpha-beta domain. Moreover, multiple types of perturbations mechanism such as multi-tone and single tone are studied. Also, researchers appplied different types of perturbations shapes such as step, ramp, chirp. Finally explanation of frequency scanning results varies a lot throughout the researches. As a result, it is essential to categoriez all of them properly and put together a comparisons between different methods to identify the suitiable method for a specific applications. Therefore, this paper addresses these issues. Hence, contirbutions of this paper goes to 
    % \begin{enumerate}
    %     \item A tabular and pictorial summarization of recent SSR events in power system 
    %     \item Analysis and comparisons of different stabality analysis methods 
    %     \item A comparitive study of different frequency scanning methods and techniques addressed in literature. 
    % \end{enumerate}

    \begin{figure*}[!t]
        \centering
        \begin{minipage}{0.32\textwidth}
            \centering
            \includegraphics[width=1\textwidth]{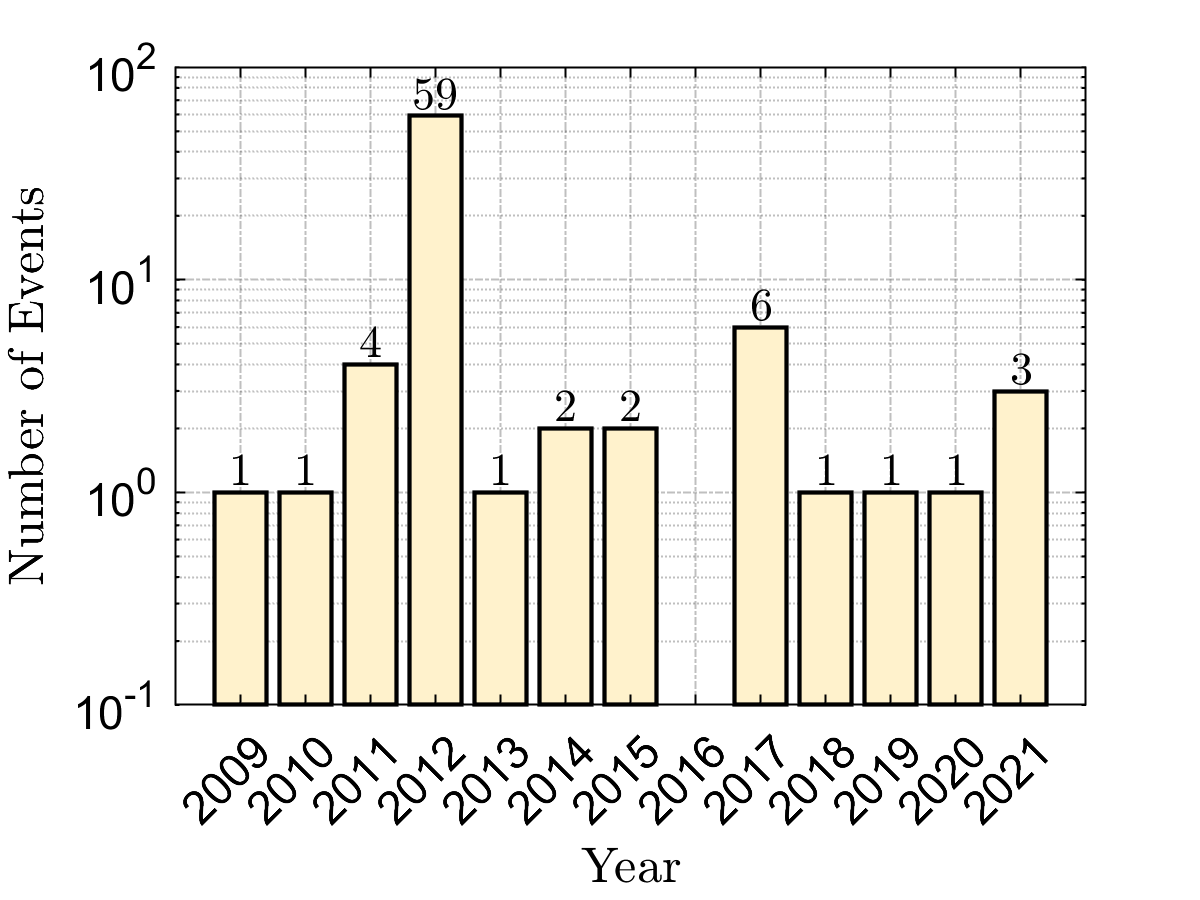}
            \caption{oscillation events per year}
            \label{fig:oscillation_events_chart}
        \end{minipage}\hfill
        \begin{minipage}{0.32\textwidth}
            \centering
            \includegraphics[width=1\textwidth]{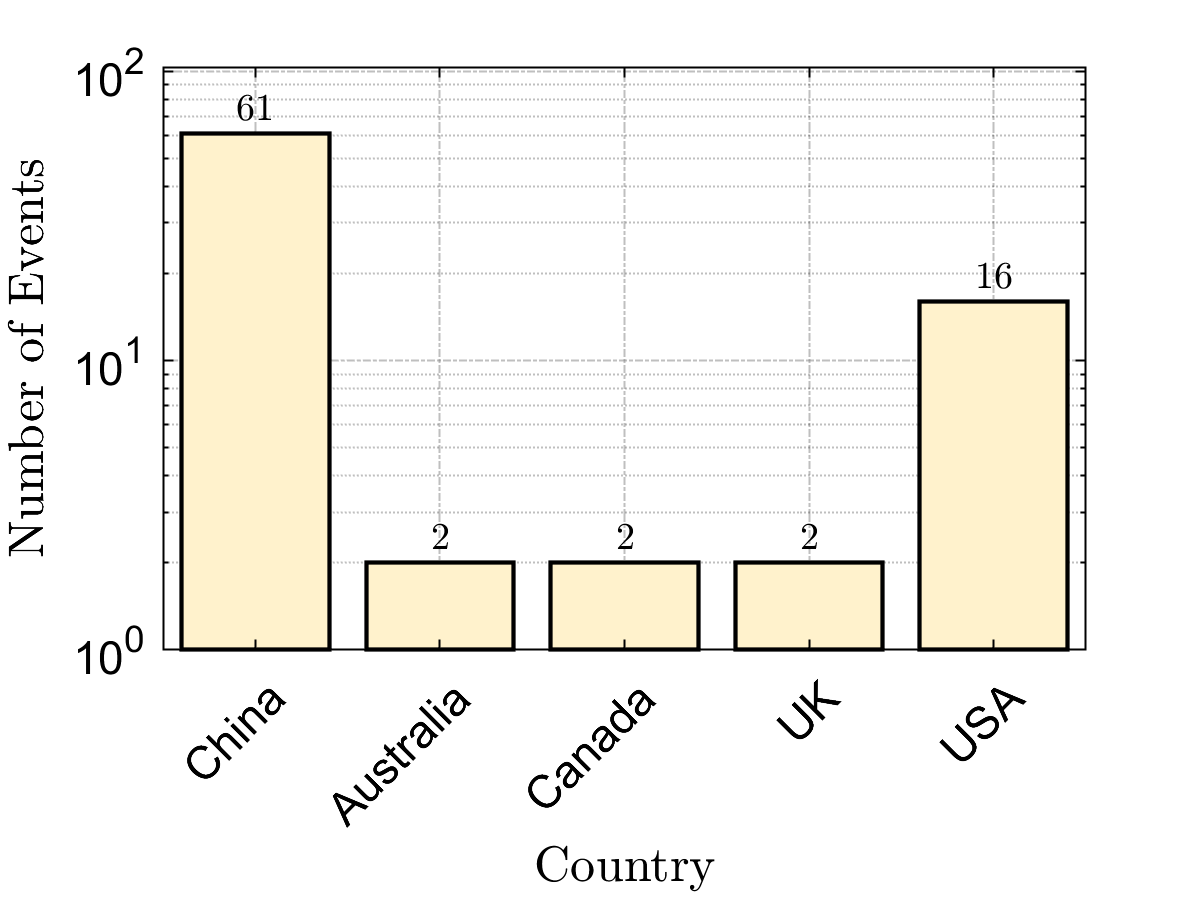}
            \caption{Oscillation events by country}
            \label{fig:oscillation_events_by_country}
        \end{minipage}\hfill
        \begin{minipage}{0.32\textwidth}
            \centering
            \includegraphics[width=1\textwidth]{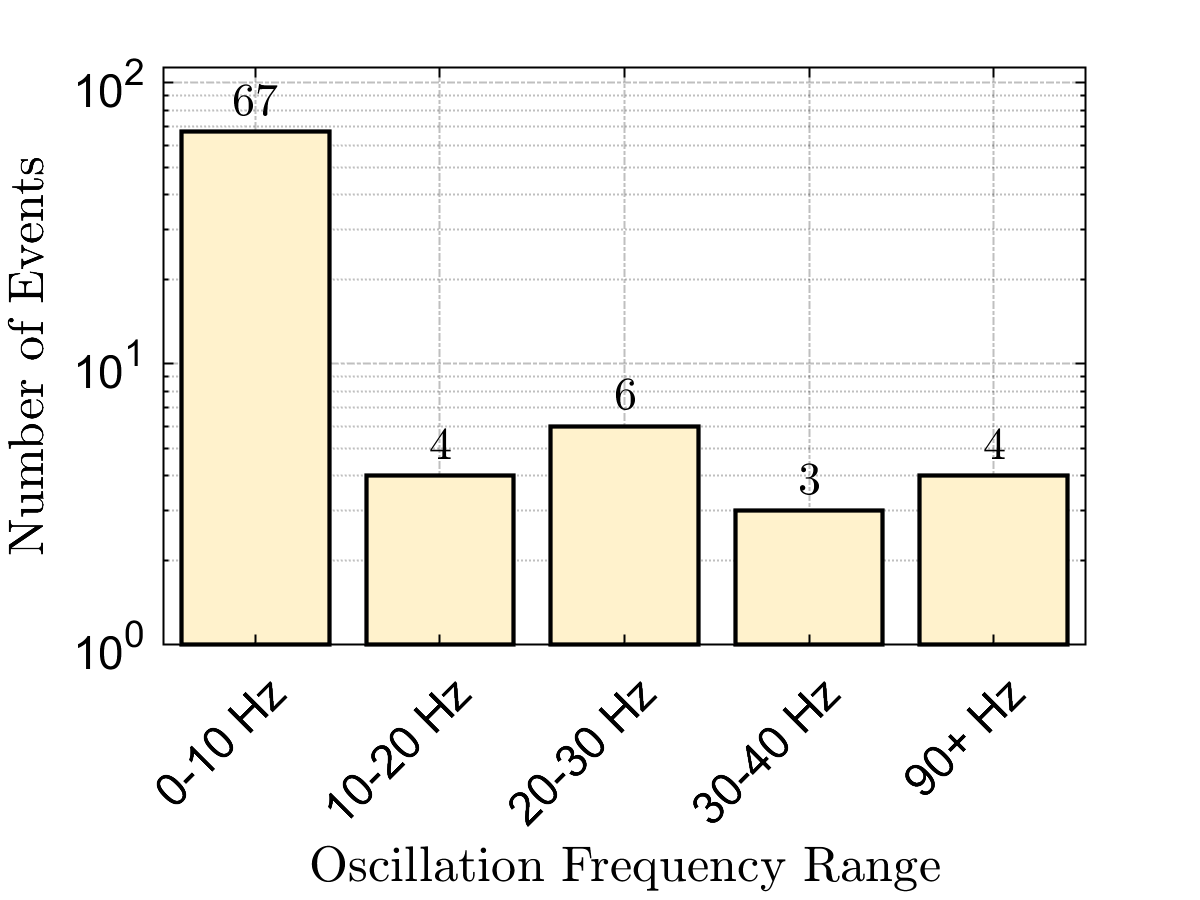}
            \caption{Oscillation events by frequency}
            \label{fig:oscillation_events_by_frequency}
        \end{minipage}
    \end{figure*}

    Frequency scanning is a prevalent tool for analyzing power system stability, with researchers applying this method across various domains such as positive sequences, dq-frame, and alpha-beta domains. Additionally, different types of perturbation mechanisms, including multi-tone and single-tone, have been studied. Researchers have also explored various perturbation shapes such as step, ramp, and chirp. However, the explanation of frequency scanning results varies significantly throughout the research. Consequently, it is essential to categorize these methods systematically and compare them to identify the most suitable approach for specific applications. Therefore, this paper addresses these issues by providing the following contributions:

    \begin{enumerate}
        % \item A comprehensive tabular and pictorial summarization of recent Subsynchronous Resonance (SSR) events in power systems.
        % \item An in-depth analysis and comparison of different stability analysis methods.
        % \item A comparative study of various frequency scanning methods and techniques discussed in the literature.
        
        \item Provides a structured historical overview of SSR/SSO incidents linked to high IBR penetration, offering insights into emerging trends.  
        \item Unifies and categorizes existing stability analysis methods (white-, black-, and gray-box) for more informed method selection.  
        \item Systematically compares frequency scanning techniques and perturbation strategies, clarifying their advantages, limitations, and applications.  
        \item Delivers practical guidance for interpreting frequency scan results to identify instability risks and inform mitigation strategies.
    \end{enumerate}

Finally, the paper is organized as follows: Section \ref{test1} reviews the history of subsynchronous oscillations, Section \ref{test2} classifies IBR stability analysis methods, Section \ref{test3} provides a detailed examination of frequency scanning techniques, Section \ref{discussion} presents a broader discussion of frequency scanning insights, and Section \ref{conclusion} concludes with the final remarks of this survey.

\begin{table*}[t]
    \centering
    \caption{\textbf{History of SSO events}}
    \begin{tabular}{c p{1.5cm} p{2.5cm} p{6cm} p{2cm} p{1.5cm}}
    \hline
    \textbf{No.} & \textbf{Year(s)} & \textbf{Location \newline or Entity} & \textbf{Event Description} & \textbf{SSO Freq(Hz)} & \textbf{Ref} \\
    \hline
    1 & 2007 & South Central Minnesota & Type-3 WPP left radially connected to 345kV series compensated line & 9.44 & \cite{ieeepesWindEnergySystems2020}, \cite{mulawarmanDetectionUndampedSubSynchronous2011a} \\
    \hline
    2 & 2009 & South Texas &  Transmission line tripping left several type-3 WPPs connected to a 345-kV series-compensated line. & 20-30 &\cite{ieeepesWindEnergySystems2020}, \cite{irwinSubsynchronousControlInteractions2011} \\
    \hline
    3 & 2010 & Oklahoma OG\&E & Two wind farms output above 80\% triggered oscillation & 13 & \cite{nercReliabilityGuidelineForced2017}\\
    \hline
    4 & 2011 & Texas & Transmission line trip induced oscillation at type-4 WPP & 4 & \cite{huangVoltageControlChallenges2012a} \\
    \hline
    5 & 2011-2014 & Oregon (BPA) & Oscillations observed during high wind generation. 5-Hz and 14-Hz oscillations were detected in different periods. & 5, 14 & \cite{nercReliabilityGuidelineForced2017} \\
    \hline
    6 & 2011-2012 & Oklahoma OG\&E & Two 3Hz wind oscillation events were triggered due to line outages in 2011 and 2012. & 3 & \cite{nercReliabilityGuidelineForced2017} \\
    \hline
    7 & 2012-2013 & North China & 58+ oscillation events from type-3 WPPs' interaction with 500-kV double circuit lines. & 6-9 & \cite{IEEEPESWindSSO2020}, \cite{xieCharacteristicAnalysisSubsynchronous2017a} \\
    \hline
    8 & 2014-2015 & Xinjiang China & 30-Hz oscillations due to type-4 WPP interaction with 750kV weak grid system& 30 & \cite{liuSubsynchronousInteractionDirectDrive2017a} \\
    \hline
    9 & 2015 & Hydro One, Canada & After the energizing of a 30-Mvar shunt capacitor at the substation. & 20(Vrms), 80 (Iac) & \cite{liUnstableOperationPhotovoltaic2018} \\
    \hline
    10 & 2016 & AEP footprint & PMUs captured oscillations for multiple days at a solar farm. & Not specified & \cite{nercReliabilityGuidelineForced2017} \\
    \hline
    11 & 2017 & Northwest China & 37-Hz and 63-Hz oscillations observed at a 600-MW type-3 WPP connected to a 220-kV weak grid & 37, 63 & \cite{vietoBehaviorModelingDamping2018}, \cite{sunDevelopmentApplicationTypeIII2020} \\
    \hline
    12 & 2017 & First Solar's solar farm, California & 7-Hz oscillations in real power, reactive power, and RMS voltage. & 7 & \cite{morjariaDeployingUtilityScalePV} \\
    \hline
    13 & 2017 & South Texas & Three separate SSO events occurred due to control software bug & 22-26 & \cite{ieeepesWindEnergySystems2020} \\
    \hline
    14 & 2015-2019 & Australia's West Murray zone & 7-Hz voltage oscillations due to low system strength and high IBR penetrations. & 7 & \cite{shahIdentifyingPotential2024}, \cite{jalaliSystemStrengthChallenges2021} \\
    \hline
    15 & 2018-2019 & Hydro One & 3.5-Hz oscillations observed in real power and reactive power for planned 230kV outage which reduced system strength of type-4 WPP & 3.5 & \cite{liAssetConditionAnomaly2019} \\
    \hline
    16 & 2019 & Great Britain (GB) & Weak grid trigger 9Hz oscillation 10 minutes before 2019 GB power system disruption and deload 800MW WPP & 9 & \cite{GBPowerSystem} \\
    \hline
    17 & 2020 & West Murray zone, Australia & 17-19 Hz voltage oscillations reported. & 17-19 & \cite{AEMOSystemStrength} \\
    \hline
    18 & 2021 & Dominion Energy, eastern U.S. & 22-Hz RMS voltage oscillations linked to a solar PV farm. 8-Hz and 82-Hz components were noted in currents and voltages.& 22, 38, 82 & \cite{wangIdentifyingOscillationsInjected2022} \\
    \hline
    19 & 2021 & Scotland & 8-Hz oscillations in RMS voltage observed on August 24, 2021. The system has high wind penetration.  & 8 & \cite{GPSTESIGWebinar} \\
    \hline
    20 & 2021 & Kaua‘i’s power system & IBR induced system-wide 18-20Hz Oscillation & 18-20 &  \cite{dongAnalysisNovember212023} \\
    \hline
    \multicolumn{6}{l}{\parbox{0.9\textwidth}{\textit{Acronyms: OG\&E - Oklahoma Gas \& Electric; BPA - Bonneville Power Administration; WF - Wind Farm; WPP - Wind Power Plant.}}}
    \end{tabular}
    \label{tab:SSCI_history}
\end{table*}

\section{Analysis of Recent SSO Events}\label{test1}

    As the penetration of renewable energy sources continues to grow, ongoing efforts to study and mitigate SSO will be crucial to ensure the stability and reliability of power systems worldwide. After reviewing literature \cite{ieeepesWindEnergySystems2020}, and IEEE PES TR-80 wind SSO task force report, all of these incidents are summarized in the Table \ref{tab:SSCI_history} with the proper reference. Over the past two decades, the power industry has observed a significant increase in SSO events, particularly associated with wind and solar power generation. These events, spanning from 2007 to 2021, have been documented across various global regions, including the United States, Canada, China, Australia, Great Britain, and Scotland, underscoring the universal challenge SSO presents to power systems integrating renewable energy sources. Table \ref{tab:SSCI_history} provides a detailed overview of these incidents, highlighting the location, event description, and oscillation frequency. The review of oscillation events from 2007 to 2021 highlights significant temporal, geographical, and frequency-specific trends depicted in Figs.~\ref{fig:oscillation_events_chart}--\ref{fig:oscillation_events_by_frequency}.
    
    % The review of oscillation events from 2007 to 2021 highlights significant temporal, geographical, and frequency-specific trends depicted in Figs.~\ref{fig:oscillation_events_chart}--\ref{fig:oscillation_events_by_frequency}.

    Fig. \ref{fig:oscillation_events_chart} reveals that oscillation events are not evenly distributed over the years, with notable spikes such as the 59 events recorded in 2013. These peaks may correspond to systemic issues or technological transitions, such as large-scale wind and solar energy integration, which introduced new challenges in grid stability. Conversely, years with fewer events may reflect improvements in grid control systems or a lower intensity of renewable energy deployment.

    Geographically, the distribution of events depicted in Fig. \ref{fig:oscillation_events_by_country} emphasizes the influence of grid size, renewable energy adoption, and monitoring practices. China reports the highest number of SSO incidents (61), followed by the USA (16), while other countries, such as Australia, Canada, and the UK, report significantly fewer events. This disparity can be attributed to the extensive adoption of renewable energy technologies, such as wind farms, in China and the USA, as well as the corresponding grid complexities. Furthermore, the dominance of China in reported events highlights the challenges posed by high renewable energy penetration in large-scale interconnected power systems.

    In terms of frequency presented in Fig. \ref{fig:oscillation_events_by_frequency}, the majority of oscillation events (67) occur within the 0--10 Hz range and often involve the inertial frequencies of synchronous generators. These low-frequency oscillations are commonly associated with interactions between renewable energy resources, weak grid conditions, and series-compensated transmission lines. Higher-frequency oscillations are becoming more prevalent, largely due to interactions between IBR controllers and the power grid. The characteristics of the bulk power grid can strongly influence these subsynchronous interactions, particularly in weak-grid situations or in cases where series or shunt capacitances create a grid resonance within this frequency range. We also anticipate that as renewable integration deepens, the overall spectrum of oscillation frequencies may shift upward, warranting further investigation.

    Many occurrences were closely tied to weak grid conditions, high penetrations of Inverter-Based Resources (IBRs), and the presence of series-compensated lines. The oscillation frequencies recorded during these events varied widely, ranging from 3 Hz oscillation to several incidents exhibiting multiple frequency components. Several regions experienced repeated SSO events. For example, Oklahoma OG\&E, South Texas, and the West Murray zone in Australia faced persistent issues. Notably, the West Murray zone dealt with prolonged challenges over four years (2015-2019), during which 7-Hz voltage oscillations were a persistent issue. In 2019, the Great Britain power system experienced a significant disruption, preceded by a 9 Hz oscillation, highlighting the potential severity of these events. In another notable incident in 2021, a solar PV farm at Dominion Energy in the eastern U.S. was linked to 22 Hz RMS voltage oscillations, showcasing the relevance of SSO issues in solar energy\cite{wangIdentifyingOscillationsInjected2022}.

    The triggering conditions for these SSO events were diverse, ranging from transmission line trips, high wind generation, and control software bugs to general grid weaknesses. This diversity highlights the complexity of SSO phenomena and the necessity for a multifaceted approach in analysis and mitigation.

    % A review of oscillation events from 2009 to 2021 highlights significant temporal, geographical, and frequency-specific trends. The temporal analysis depicted in Fig. \ref{fig:oscillation-events-chart} reveals that oscillation events are not evenly distributed over the years, with notable spikes such as the 59 events recorded in 2013. These peaks may correspond to systemic issues or technological transitions, such as large-scale wind and solar energy integration, which introduced new challenges in grid stability. Conversely, years with fewer events may reflect improvements in grid control systems or a lower intensity of renewable energy deployment.

    % \begin{figure} 
    %     \centering 
    %     \includegraphics[width=0.5\textwidth]{oscillation_events_chart.png} 
    %     \caption{Placeholder} 
    %     \label{fig:referenceLink} 
    % \end{figure}
    
    % \begin{figure} 
    %     \centering 
    %     \includegraphics[width=0.5\textwidth]{oscillation_events_by_country.png} 
    %     \caption{Placeholder} 
    %     \label{fig:referenceLink} 
    % \end{figure}
    
    % \begin{figure} 
    %     \centering 
    %     \includegraphics[width=0.5\textwidth]{oscillation_events_by_frequency.png} 
    %     \caption{Placeholder} 
    %     \label{fig:referenceLink} 
    % \end{figure}

    \begin{table*}[!tbh]
        \centering
        \caption{Classification of Different Approaches of Stabality Analysis}
        \label{tab:stability_analysis_review}
        \begin{tabular}{l>{\raggedright}p{3cm}>{\raggedright}p{4cm}>{\raggedright}p{4cm}>{\raggedright\arraybackslash}p{3cm}}
          \toprule
          \textbf{Type} & \textbf{Method} & \textbf{Pros} & \textbf{Cons} & \textbf{Ref} \\
          \midrule
          \multirow{4}{*}{White-box} & State-space & Detailed insights into system dynamics & Requires system knowledge & \cite{moharanaSSRMitigationWind2012} \cite{moharanaSSRAlleviationSTATCOM2014} \cite{fanModelingDFIGBasedWind2010} \cite{xuSmallSignalStabilityAnalysis2020} \cite{trevisanAnalysisLowFrequency} \cite{aminSmallSignalStabilityAssessment2017} \cite{suriyaarachchiProcedureStudySubSynchronous2013} \cite{raumaResonanceAnalysisWind2012} \cite{liWindWeakGrids2020} \\
          \cline{2-5}
          & Transfer function based & Suitable for linear systems & Not ideal for non-linear systems & \cite{bajracharyaUnderstandingTuningTechniques2008} \cite{alawasaModelingAnalysisSuppression2013} \cite{alawasaActiveMitigationSubsynchronous2014} \cite{hailianxieMitigationSSRPresence2014} \cite{sainzAssessmentSubsynchronousOscillations2019} \\
          \cline{2-5}
          & Impedance Model & Provides frequency-domain insights & Challenging for complex systems & \cite{fanNyquistStabilityCriterionBasedSSRExplanation2012} \cite{chernetOnlineVariationWind2016} \cite{alawasaActiveMitigationSubsynchronous2014} \cite{aminSmallSignalStabilityAssessment2017} \cite{aminUnderstandingOriginOscillatory2017} \cite{liuSubsynchronousInteractionDirectDrive2017a} \\
          \cline{2-5}
          & EMT Simulation & Accurate time-domain simulation & Computationally intensive & \cite{hailianxieMitigationSSRPresence2014} \cite{liuSubsynchronousInteractionDirectDrive2017a} \cite{liReplicatingRealWorldWind2020a} \\
          \midrule
          \multirow{4}{*}{Black-box} & Frequency scanning & Does not require system knowledge & Limited to frequency domain & \cite{elfayoumyComprehensiveApproachSubsynchronous2003} \cite{nathStudySubSynchronousControl2012} \cite{sahniAdvancedScreeningTechniques2012} \cite{badrzadehGeneralMethodologyAnalysis2013} \cite{chengReactanceScanCrossoverBased2013} \cite{guptaFrequencyScanningStudy2013} \cite{wenStabilityAnalysisThreephase2014} \cite{renRefinedFrequencyScan2016} \cite{karaagacSafeOperationDFIGBased2018} \cite{liaoGeneralRulesUsing2018} \cite{ryggApparentImpedanceAnalysis2017} \cite{trevisanAnalyticallyValidatedSSCI2021} \cite{fanIdentifyingDQDomainAdmittance2021} \cite{lwinFrequencyScanConsiderations2019} \cite{liuAnalysisDesignImplementation2020} \cite{matsuoOptimizedFrequencyScanning2020} \cite{jacobsComparativeStudyFrequency2023} \cite{mengNewSequenceDomain2023} \cite{ramakrishnaDQAdmittanceExtraction2023} \cite{shirinzadFrequencyScanBased} \\
          \cline{2-5}
          & Vector Fitting Method & Robust for high-order systems & Can be complex to implement & \cite{bakhshizadehImprovingImpedanceBasedStability2018} \\
          \cline{2-5}
          & Eigen System Realization & Effective for modal analysis & Requires data preprocessing & \cite{fanTimeDomainMeasurementBasedDQFrame2021} \\
          \cline{2-5}
          & Dissipating Energy Flow & Insightful for energy distribution & Limited by measurement accuracy & \cite{dongAnalysisNovember212023} \cite{maslennikovDissipatingEnergyFlow2017} \cite{chenEnergybasedMethodLocation2013} \\
          \midrule
          Grey-box & Combines model knowledge and data-driven methods & Requires both system knowledge and data & - & \cite{zhuParticipationAnalysisImpedance2022} \cite{aminGrayBoxMethodStability2019} \\
          \bottomrule
        \end{tabular}
      \end{table*}
    
      \begin{figure}[!b] 
        \centering 
        \includegraphics[width=0.5\textwidth]{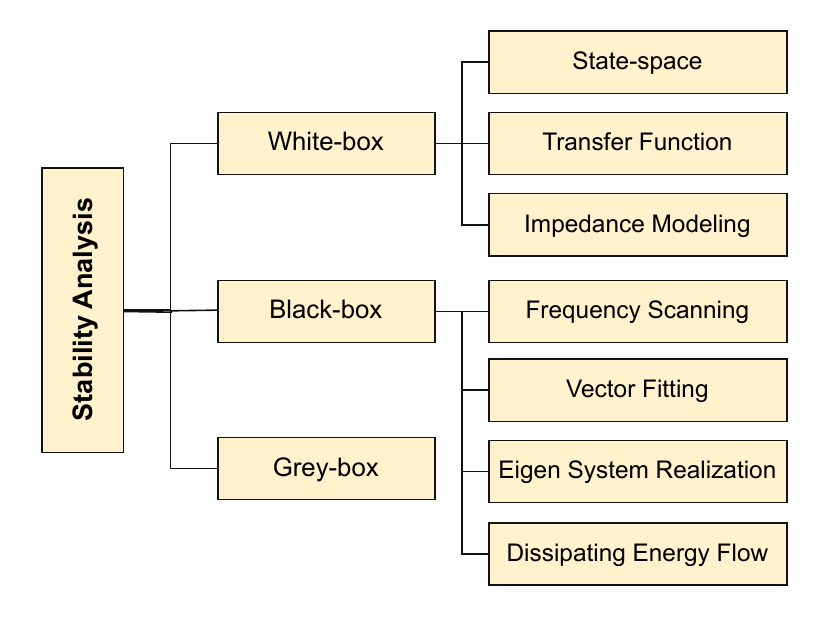} 
        \caption{Classificaiton of IBR Stability Analysis} 
        \label{fig:stabality-analysis-classification} 
    \end{figure}

\section{Stabality Analysis}\label{test2}
Identifying the possible oscillation is the key factor of stable grid operation. Investigating the origins of oscillations associated with IBR presents a multifaceted challenge. This complexity is attributed to the disparate IBR controller designs sourced from numerous vendors, coupled with the restricted access to, or the proprietary nature of, well-validated IBR models \cite{dongAnalysisNovember212023}.

Existing literature explored different approaches to identify the oscillation. In this papers, all of those approaches are categorized under white-box, black-box, and gray-box. After carefully reviewing, a comprehensive categorization is presented in Fig. \ref{fig:stabality-analysis-classification}. The white-box approach considers that the whole system is known, containing physical and control parameters. White-box stability analysis is categorized down into eigenvalue analysis using state-space modeling \cite{moharanaSSRAlleviationSTATCOM2014, fanModelingDFIGBasedWind2010, xuSmallSignalStabilityAnalysis2020, trevisanAnalysisLowFrequency, aminSmallSignalStabilityAssessment2017, suriyaarachchiProcedureStudySubSynchronous2013, raumaResonanceAnalysisWind2012, liWindWeakGrids2020}
, transfer function based analysis \cite{bajracharyaUnderstandingTuningTechniques2008, alawasaModelingAnalysisSuppression2013, alawasaActiveMitigationSubsynchronous2014, hailianxieMitigationSSRPresence2014, sainzAssessmentSubsynchronousOscillations2019}
, impedance model-based analysis \cite{fanNyquistStabilityCriterionBasedSSRExplanation2012}
\cite{chernetOnlineVariationWind2016}
\cite{alawasaActiveMitigationSubsynchronous2014}
\cite{aminSmallSignalStabilityAssessment2017}
\cite{aminUnderstandingOriginOscillatory2017}
\cite{liuSubsynchronousInteractionDirectDrive2017a} and EMT simulation
\cite{hailianxieMitigationSSRPresence2014, liuSubsynchronousInteractionDirectDrive2017a, liReplicatingRealWorldWind2020a}. On the other hand, black-box IBR models are provided in binary format by the manufacturer, which can be solved in time-domain simulation and represent non-linear dynamics with high fidelity. Different literature analyzed the black-box model's sub-synchronous oscillation in different ways. These methods includes frequency scanning \cite{elfayoumyComprehensiveApproachSubsynchronous2003, nathStudySubSynchronousControl2012, sahniAdvancedScreeningTechniques2012, badrzadehGeneralMethodologyAnalysis2013, chengReactanceScanCrossoverBased2013, guptaFrequencyScanningStudy2013, wenStabilityAnalysisThreephase2014, renRefinedFrequencyScan2016, karaagacSafeOperationDFIGBased2018, liaoGeneralRulesUsing2018, ryggApparentImpedanceAnalysis2017, trevisanAnalyticallyValidatedSSCI2021, fanIdentifyingDQDomainAdmittance2021, lwinFrequencyScanConsiderations2019, liuAnalysisDesignImplementation2020, matsuoOptimizedFrequencyScanning2020, jacobsComparativeStudyFrequency2023, mengNewSequenceDomain2023, ramakrishnaDQAdmittanceExtraction2023, shirinzadFrequencyScanBased}, Vector Fitting (VF) method \cite{bakhshizadehImprovingImpedanceBasedStability2018}, Eigen System Realization \cite{fanTimeDomainMeasurementBasedDQFrame2021} and Dissipating Energy Flow \cite{dongAnalysisNovember212023, maslennikovDissipatingEnergyFlow2017, chenEnergybasedMethodLocation2013}.

The system operator prefers white-box IBR models for stability analysis. White-box models are state-space equations containing all of the physical control and system states. The state equation yields eigenvectors and the participation factor, which indicates the responsible state for oscillation. Therefore, to bridge the gap between black-box models by manufacturer and white-box model desired by the system operator, multiple literatures \cite{zhuParticipationAnalysisImpedance2022, aminGrayBoxMethodStability2019} put an effort into building a gray-box approach that connects the gap between two methods.

% Fig. \ref{fig:stabality-analysis-classification} summarizes the stability analysis approach with literature references. Two methods are commonly used for IBR stability analysis \cite{jacobsComparativeStudyFrequency2023}, i.e., eigenvalue analysis using state-space model, which is considered as white box modeling and impedance scanning method using bode plot or Nyquist plot. Both ways have their merits and demerits. The state-space-based method requires a detailed model of the converter \cite{jacobsComparativeStudyFrequency2023}, which is not always possible for a proprietary IBR-based system. Moreover, a model must be reformulated for each structural change. Since the power electronics converter is highly non-linear, a model must be linearized around an equilibrium point. On the other hand, the frequency scanning method is a time-intensive method, and it is prone to noise. The literature survey indicates that state-space and frequency scanning methods are mostly utilized for stability analysis. The grey-box method is the combination of both approaches. Therefore, In the subsequent section, frequency scanning methods will be discussed and categorized in detail.

Fig. \ref{fig:stabality-analysis-classification} summarizes the primary stability assessment approaches and their supporting literature. Two frequently employed techniques for IBR stability analysis are eigenvalue-based state-space methods (white-box) and impedance-based frequency scanning methods (Bode or Nyquist plots). Each approach has inherent trade-offs: state-space modeling demands detailed, open-form converter models-which may not be available for proprietary systems-and requires re-linearization for every structural change, while frequency scanning is computationally intensive and sensitive to measurement noise. Given the nonlinear nature of power electronic converters, white-box modeling often involves complex linearization around specific operating points. Current literature shows these two methods dominate practical application, although hybrid (gray-box) approaches are emerging to bridge their respective limitations. The following section provides a detailed categorization of frequency scanning techniques.

\section{Frequency Scanning and Stabality Analysis}\label{test3}

\begin{figure*}
	\centering 
	\includegraphics[width=0.85\textwidth]{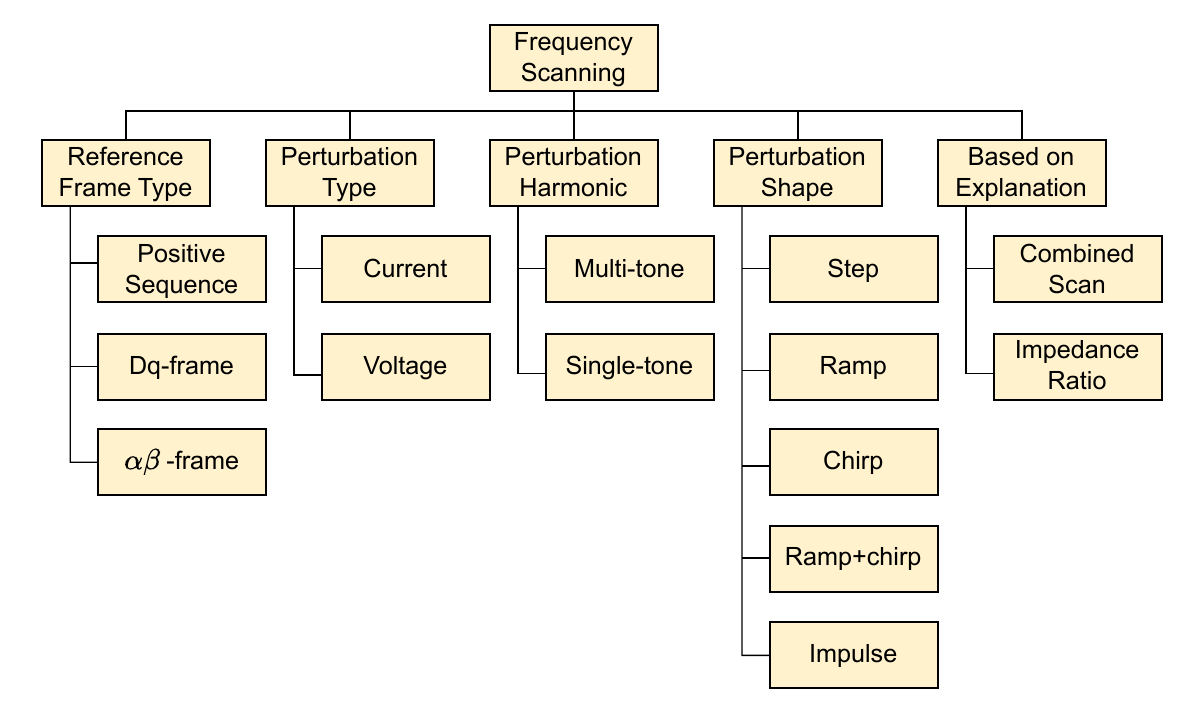} 
	\caption{Frequency scanning classifications and different methods} 
	\label{fig:fscan-classification} 
\end{figure*}

    Frequency scanning proved to be a prevalent tool for characterizing IBR and power systems. Existing literature \cite{elfayoumyComprehensiveApproachSubsynchronous2003, nathStudySubSynchronousControl2012, sahniAdvancedScreeningTechniques2012, badrzadehGeneralMethodologyAnalysis2013, chengReactanceScanCrossoverBased2013, guptaFrequencyScanningStudy2013, wenStabilityAnalysisThreephase2014, renRefinedFrequencyScan2016, karaagacSafeOperationDFIGBased2018, liaoGeneralRulesUsing2018, ryggApparentImpedanceAnalysis2017, trevisanAnalyticallyValidatedSSCI2021, fanIdentifyingDQDomainAdmittance2021, lwinFrequencyScanConsiderations2019, liuAnalysisDesignImplementation2020, matsuoOptimizedFrequencyScanning2020, jacobsComparativeStudyFrequency2023, mengNewSequenceDomain2023, ramakrishnaDQAdmittanceExtraction2023, shirinzadFrequencyScanBased} explored mainly three types of frequency scanning techniques such as (1) positive sequence (ps), (2) dq-frame, and (3) $\alpha\beta$-frame scanning . Ref. \cite{jacobsComparativeStudyFrequency2023} shortly explained three types of scannings and their relative advantage and computation complexities and feasibility in characterizing IBRs and generators. The dimensionality of the frequency-dependent model obtained varies based on the reference frame selected. The positive-sequence impedance model is a Single-Input Single-Output (SISO) system, while both \( dq \) and \( \alpha \beta \) impedance models are Multiple-Input Multiple-Output (MIMO) systems. Nonetheless, the scanning procedures for these three methods are identical. However, extracting a SISO impedance can be accomplished with a single scan. In contrast, a MIMO impedance necessitates separate perturbations for each axis, such as the d and q-axis separately \cite{jacobsComparativeStudyFrequency2023}. Ref.  \cite{jacobsComparativeStudyFrequency2023} also suggested not all types of frequency scans are suitable for a specific application. Such as the ref. \cite{jacobsComparativeStudyFrequency2023} suggested that the dq-scan provides the highest accuracy for full-size converter-based wind power systems (type-4).
    Fig. \ref{fig:fscan-classification} depicts the classification of frequency based on recent literature.  The classification is presented based on different categories. In the subsequent sections, the categorization will be explained briefly. 

    \subsection{Frequency Scanning Mechanism}

    The fundamental steps of the frequency scanning procedure remain consistent regardless of the chosen reference frame (positive sequence, dq-frame, or \(\alpha\beta\)-frame), perturbation type (voltage or current), perturbation shape (step, ramp, chirp), or scanning approach (single-tone or multi-tone). Figs. \ref{fig:ps-frequency-scan}, \ref{fig:cs}, and \ref{fig:vs} illustrate examples of positive-sequence scanning, voltage-source perturbation, and current-source perturbation, respectively. Additional details on single-tone and multi-tone scanning techniques can be found in Subsections \ref{single-tone-frequency-scanning} and \ref{multi-tone-frequency-scan}. The general procedure, as widely adopted in literature (e.g., \cite{renRefinedFrequencyScan2016}), is summarized as follows:

    \begin{enumerate}
        \item \textbf{Establish Steady-State Conditions:}  
        Begin by running the system under its normal operating point until it reaches steady state. It is crucial that the system operates in a linear region around this point to ensure the validity of small-signal perturbation and linearization-based methods.

        \item \textbf{Inject Perturbation:}  
        Superimpose a small, sinusoidal perturbation of known frequency on the system’s voltage or current. The perturbation amplitude must be large enough to provide a good signal-to-noise ratio, yet small enough to avoid driving the system into non-linear regions. Record the resulting voltage and current waveforms at the point of interest.

        \item \textbf{Frequency Domain Analysis:}  
        Apply a Fast Fourier Transform (FFT) to the measured signals to extract voltage and current phasors at the injection frequency. The corresponding impedance at that frequency is then computed as \( Z(f) = V(f)/I(f) \). The detailed steps for FFT-based amplitude extraction are presented in Algorithm \ref{algo:fft-algorithm}.

        \item \textbf{Frequency Sweep:}  
        Repeat the perturbation and measurement process over a range of frequencies to build a full impedance profile. For example, to cover 0--120 Hz at a resolution of 0.5 Hz, systematically inject perturbations at each discrete frequency and record the resulting voltage abd current and post-process for impedance.

    \end{enumerate}
    \noindent
    This iterative, frequency-by-frequency scanning procedure ultimately yields a frequency-dependent impedance model of the subsystem, enabling stability assessments and resonance analysis.

    % \begin{algorithm}[h]
    %     \caption{FFT-Based Amplitude Extraction}
    %     \begin{algorithmic}[1]
    %     \Require Sampling frequency \(F_s\), number of samples \(N\), FFT result \(X(k)\)
    %     \Ensure Frequency vector and single-sided amplitude spectrum
        
    %     \Statex \textbf{Step 1: Amplitude Calculation}
    %     \begin{itemize}
    %         \item Compute amplitude from FFT bin: \(|X(f_k)| = |X(k)|/N\)
    %     \end{itemize}
        
    %     \Statex \textbf{Step 2: Construct Single-Sided Spectrum}
    %     \begin{itemize}
    %         \item Keep only the first \(N/2\) points (positive frequencies).
    %         \item Multiply amplitudes (except the DC component) by 2 to account for discarded negative frequencies.
    %     \end{itemize}
        
    %     \Statex \textbf{Step 3: Frequency Axis Construction}
    %     \begin{itemize}
    %         \item Frequency resolution: \(\Delta f = F_s/N\)
    %         \item Frequency vector: \([0, \Delta f, 2\Delta f, \ldots, (N/2-1)\Delta f]\)
    %     \end{itemize}
        
    %     \Statex \textbf{Output:} Frequency vector and corresponding amplitude spectrum.
    %     \end{algorithmic}
    %     \label{algo:fft-algorithm}
    % \end{algorithm}

    \begin{algorithm}[h]
        \caption{FFT Interpretation for Amplitude Extraction}
        \begin{algorithmic}[1]
        \Require Sampling frequency $F_s$, Number of samples $N$, FFT result $X(k)$
        \Ensure Frequency vector and Amplitude spectrum
        
        \State \textbf{Step 1: Get Amplitude from FFT Bin}
        \begin{itemize}
            \item Calculate amplitude: $|X(f_k)| = |X(k/N)|$
        \end{itemize}
        
        \State \textbf{Step 2: Discard Negative Part of Spectrum}
        \begin{itemize}
            \item Identify two-sided spectrum
            \item Discard the second half of the array
            \item Multiply remaining points by 2, except DC at 0Hz
        \end{itemize}
        
        \State \textbf{Step 3: Convert Bin to Frequency}
        \begin{itemize}
            \item Calculate total number of frequency points: $\frac{N}{2}$
            \item Calculate frequency interval: $\Delta f = \frac{F_s}{N}$
            \item Construct frequency vector: $[0:\Delta f:(\frac{F_s}{2} - \Delta f)]$
        \end{itemize}
        
        \State \textbf{Output: Frequency vector and Single-sided Amplitude spectrum}
        
        \end{algorithmic}
        \label{algo:fft-algorithm}
    \end{algorithm}

    \subsection {Frequency Scanning: Reference Frame Type}
    In frequency scanning, three main types of reference frames are commonly used to analyze system behavior: positive sequence frequency scanning, dq-frame frequency scanning, and $\alpha\beta$-frame frequency scanning.
    \subsubsection{\textbf{Positive Sequence Scanning}}

    Positive sequence scanning (ps-scan) is a crucial technique in power systems analysis, especially for studying the impedance characteristics of IBR connected to weak grids or compensated networks~\cite{karaagacSafeOperationDFIGBased2018}. The positive sequence impedance model is a Single-Input Single-Output (SISO) system, providing a simplified representation of the system's behavior at the perturbation frequency. 
    
\begin{figure} [!t]
        \centering 
        \includegraphics[width=0.5\textwidth]{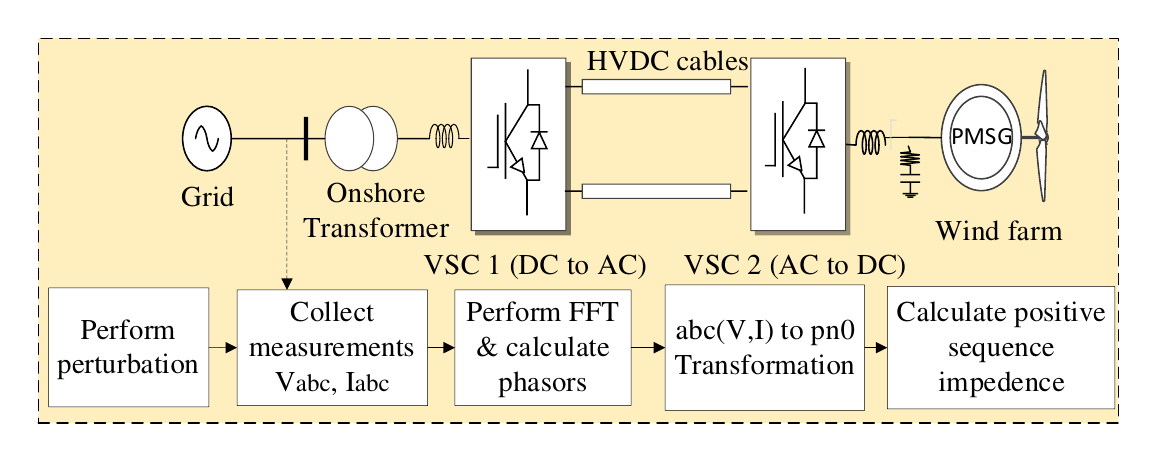} 
        \caption{Procedure of performing positive sequence frequency scanning} 
        \label{fig:ps-frequency-scan} 
    \end{figure}

    Fig. \ref{fig:ps-frequency-scan} depicts the ps-scanning mechanism for HVDC wind farm system. This method leverages the application of sinusoidal perturbation signals in the sequence domain to derive the impedance of the converter subsystem based on the positive sequence components of the voltage and current signals at the point of interconnection (POI). 
    
    % The positive-sequence scanning mechanism injects a small perturbation into the system, measures voltage and current, perform FFT and transforms the resulting phasor signals into positive-sequence components, and then calculates the frequency-dependent impedance from these phasors. 

    \vspace{0.5cm}
    The relationship between the three-phase voltages ($V_a, V_b, V_c$) and the sequence components ($V_0, V_1, V_2$) is expressed through a transformation matrix. Specifically, the three-phase voltages can be represented as:

    \begin{equation}
    \begin{pmatrix}
    V_a \\
    V_b \\
    V_c
    \end{pmatrix} =
    \begin{pmatrix}
    1 & 1 & 1 \\
    1 & a^2 & a \\
    1 & a & a^2
    \end{pmatrix}
    \begin{pmatrix}
    V_0 \\
    V_1 \\
    V_2
    \end{pmatrix}
    \end{equation}

    \noindent
    where $a = -\frac{1}{2} + \frac{\sqrt{3}}{2}i$ represents a complex operator corresponding to a 120-degree phase shift. Within this context, the positive sequence component ($V_{\text{1}}$) is calculated using the formula:

    \begin{equation}
    V_1 = \frac{1}{3}(V_a + aV_b + a^2V_c)
    \label{eq:pos_seq}
    \end{equation}

    \noindent
    This equation accurately determines the positive sequence component from the three-phase phasors, providing a reliable means to analyze the system.

    \vspace{0.5cm}
    % The ps-scan technique follows a structured process to extract impedance information illustrated in Fig. \ref{fig:ps-frequency-scan}. Initially, the scanned grid is brought to a steady-state condition. Following this, a positive-sequence voltage at the desired frequency, $V_{\text{inj}}(f)$, is injected between the equivalent source and the grid terminals. Subsequently, the voltage ($V_{\text{meas}}$) and current ($I_{\text{meas}}$) at the grid terminals are measured. The Fast Fourier Transform (FFT) is then applied to both $V_{\text{meas}}$ and $I_{\text{meas}}$ to obtain their phasor values at the frequency of the injected voltage, $V_{\text{inj}}(f)$. Then, the positive sequence component is calculated for both voltage and current using the equation \ref{eq:pos_seq}. Finally, the positive sequence impedance ($Z_{\text{meas}}(f)$) is calculated using the relation:

    The ps-scan technique follows a structured process to extract impedance information. Initially, the scanned grid is brought to a steady-state condition. Following this, a positive-sequence voltage at the desired frequency, $V_{\text{inj}}(f)$, is injected between the equivalent source and the grid terminals. Subsequently, the voltage ($V_{\text{abc}}$) and current ($I_{\text{abc}}$) at the grid terminals are measured. The Fast Fourier Transform (FFT) is then applied to both $V_{\text{abc}}$ and $I_{\text{abc}}$ to obtain their phasor values at the frequency of the injected voltage, $V_{\text{inj}}(f)$. Then, the positive sequence component is calculated for both voltage and current using the equation \ref{eq:pos_seq}. Finally, the positive sequence impedance ($Z_1(f)$) is calculated for each perturb frequency using the relation:

    \begin{equation}
    Z_1(f) = \frac{V_1(f)}{I_1(f)}
    \end{equation}

    \noindent
    In this scenario, $V_{\text{abc}}$ is considered a phase-to-ground voltage, while $I_{\text{abc}}$ represents the current measured at the same phase.

    \vspace{0.5cm}
    A significant advantage of the positive sequence scanning technique is its computation speed, which is faster than that of dq and $\alpha \beta$ scanning methods \cite{jacobsComparativeStudyFrequency2023}\cite{shahSequenceImpedanceMeasurement2022}. This efficiency makes ps-scan a preferred choice for impedance extraction in power systems, ensuring quick and accurate analysis, crucial for the safe and reliable operation of wind power plants and associated transmission systems.

    % \begin{table*}[t]
    %     \centering
    %     \caption{\textbf{Comparison between ps-scan, dq-scan, and $\alpha\beta$-scan methods}}
    %     \begin{tabular}{c p{4.5cm} p{4.5cm} p{4.5cm}}
    %     \hline
    %     \textbf{Item} & \textbf{ps-scan} & \textbf{dq-scan} & \textbf{$\alpha\beta$-scan} \\
    %     \hline
    %     \hline
    %     Stability criterion & Bode plot & Nyquist criterion & Generalized Nyquist criterion \\
    %     \hline
    %     Parameter & $Z_{ps}(f)$ and $Z_{ps}(c)(f)$ & $L(f)$ in dq-frame & $L(f)$ in $\alpha\beta$-frame \\
    %     \hline
    %     Prediction reliability & Good, depends on impedance model accuracy & Very good, captures dq coupling, depends on impedance ratio accuracy & Good, depends on eigenvalue accuracy \\
    %     \hline
    %     Stability margin & Yes, visualized & Yes & Yes \\
    %     \hline
    %     Oscillation frequency & Magnitude intersection & Intersection with the unit circle & Intersection of dominant eigenlocus with the unit circle \\
    %     \hline
    %     Computational burden & Lowest & Generally, 2 × equivalent ps-scan & 2 × equivalent ps-scan \\
    %     \hline
    %     \hline
    %     \end{tabular}
    %     \label{tab:scan_comparison}
    % \end{table*}

    \begin{table*}[!b]
        \centering
        \caption{\textbf{Comparison between ps-scan, dq-scan, and $\alpha\beta$-scan methods}}
        \begin{tabular}{c p{4.5cm} p{4.5cm} p{4.5cm}}
        \hline
        \textbf{Item} & \textbf{ps-scan} & \textbf{dq-scan} & \textbf{$\alpha\beta$-scan} \\ [1ex]
        \hline
        \hline
        Stability criterion & Bode plot & Nyquist criterion & Generalized Nyquist criterion \\ [1.5ex]
        \hline
        Parameter & $Z_{ps}(f)$ and $Z_{ps}(c)(f)$ & $L(f)$ in dq-frame & $L(f)$ in $\alpha\beta$-frame \\ [1.5ex]
        \hline  
        Prediction reliability & Good, depends on impedance model accuracy & Very good, captures dq coupling, depends on impedance ratio accuracy & Good, depends on eigenvalue accuracy \\ [1.5ex]
        \hline
        Stability margin & Yes, visualized & Yes & Yes \\ [1.5ex]
        \hline
        Oscillation frequency & Magnitude intersection & Intersection with the unit circle & Intersection of dominant eigenlocus with the unit circle \\ [1.5ex]
        \hline
        Computational burden & Lowest & Generally, 2 × equivalent ps-scan & 2 × equivalent ps-scan \\ [1.5ex]
        \hline
        \hline
        \end{tabular}
        \label{tab:scan_comparison}
    \end{table*}

    \subsubsection{\textbf{dq-frame Frequency Scanning}}
    The relationship between \( dq \)-currents and \( dq \)-voltages is represented by:
    
    \begin{equation}
    \begin{bmatrix}
    i_d(s) \\
    i_q(s)
    \end{bmatrix}
    = \boldsymbol{Y}_{dq}(s)
    \begin{bmatrix}
    v_d(s) \\
    v_q(s)
    \end{bmatrix}
    = \begin{bmatrix}
    Y_{dd}(s) & Y_{dq}(s) \\
    Y_{qd}(s) & Y_{qq}(s)
    \end{bmatrix}
    \begin{bmatrix}
    v_d(s) \\
    v_q(s)
    \end{bmatrix}
    \end{equation}
    where \( i_d(s) \) and \( i_q(s) \) are the \( dq \)-currents, \( v_d(s) \) and \( v_q(s) \) are the \( dq \)-voltages, and \( \boldsymbol{Y}_{dq}(s) \) is the \( dq \)-admittance matrix. This equation captures the coupling in the \( dq \)-frame, which sequence and \( \alpha \beta \)-scans might misrepresent in certain grid components\cite{cespedesImpedanceModelingAnalysis2014}\cite{trevisanAnalyticallyValidatedSSCI2021a}\cite{ryggModifiedSequenceDomainImpedance2016a}.
    The scanning procedure begins with the converter, where a step change is applied to the d-axis voltage, and the resulting dq-currents are measured at different frequencies. The process is repeated for the q-axis voltage. These signals are then transformed from the dq-frame back to the abc-frame using a dq-to-abc transformation and injected into the grid for further analysis. After measuring the currents and voltages in the abc-frame, the data is converted back to the dq-frame for Fourier analysis. The admittance components are computed as follows:
        \begin{equation}
        \begin{aligned}
        & Y_{dd}(f_i) = \frac{i_d^d(f_i)}{v_d^d(f_i)}, & Y_{dq}(f_i) = \frac{i_d^q(f_i)}{v_q^q(f_i)}, \\
        & Y_{qd}(f_i) = \frac{i_q^d(f_i)}{v_d^d(f_i)}, & Y_{qq}(f_i) = \frac{i_q^q(f_i)}{v_q^q(f_i)}.
        \end{aligned}
        \end{equation}
    For grid scanning, a similar current injection approach is employed, and the grid's impedance is calculated using the same methodology. The resulting impedance values for the d- and q-axes can be expressed as:
        \begin{equation}
        \begin{aligned}
        & Z_{dd}(f_i) = \frac{v_d^d(f_i)}{i_d^d(f_i)}, & Z_{dq}(f_i) = \frac{v_d^q(f_i)}{i_q^q(f_i)}, \\
        & Z_{qd}(f_i) = \frac{v_q^d(f_i)}{i_d^d(f_i)}, & Z_{qq}(f_i) = \frac{v_q^q(f_i)}{i_q^q(f_i)}.
        \end{aligned}
        \end{equation}\\
    % Frequency scanning is performed by injecting sinusoidal signals at various frequencies and recording the resulting currents. These signals are then processed using Fourier Transform to derive the frequency-dependent admittance.

    For system stability analysis, the loop gain is subsequently calculated via the Generalized Nyquist Criterion (GNC).  Specifically, the loop gain-defined as the product of the grid’s dq-impedance and the converter’s dq-admittance—is given in Equation \ref{eq: grid-CGI-L(s)-gain}.

    \subsubsection{\textbf{$\alpha\beta$-frame Frequency Scanning}}
    The $ \alpha \beta $-scan procedure is employed to deduce the small-signal transfer functions that map the input currents (or voltages) to the output voltages (or currents) within the steady $ \alpha \beta $ reference frame \cite{jacobsComparativeStudyFrequency2023}\cite{wangUnifiedImpedanceModel2018}. Aside from the transformations applied, this method is analogous to the $ d q $-scan previously explained.

    The impedance on the grid side in the $ \alpha \beta $-frame can be converted from the phasor-domain scan using the equation:
    \begin{equation}
    \mathbf{Z}_g^{\alpha \beta}(f) = \mathbf{T}_{C'} \cdot \begin{bmatrix} Z_g^{ps}(f) & 0 \\ 0 & Z_g^{ps*}(-f) \end{bmatrix} \cdot \mathbf{T}_{C'}
    \end{equation}
    where $ \mathbf{T}_{C'} = \frac{1}{\sqrt{2}} \begin{bmatrix} 1 & j \\ 1 & -j \end{bmatrix} $, and $ \mathbf{T}_{C'} $ represents the complex transformation matrix.

    Concurrently, the scan on the converter side gauges the $ \alpha \beta $ admittance of the converter, represented as:
    \begin{equation}
    \mathbf{Y}_c^{\alpha \beta}(f) = \begin{bmatrix} Y_{\alpha, \alpha}(f) & Y_{\alpha, \beta}(f) \\ Y_{\beta, \alpha}(f) & Y_{\beta, \beta}(f) \end{bmatrix}
    \end{equation}
    with $ Y_{j, k}(f) = \frac{I_j(f)}{V_k(f)} $, where $ j, k \in \{ \alpha, \beta \} $. Here, $ I $ and $ V $ denote current and voltage, respectively, and the subscripts signify the respective measured and perturbed axes.

    % \begin{table}[!ht]
    %     \centering
    %     \caption{Comparison of Different Scanning Methods}
    %     \label{tab:scanning_methods}
    %     \begin{tabular}{m{2cm} m{2cm} m{2cm} m{4cm} m{3cm}}
    %     \toprule
    %     Item & - & ps-scan & dq-scan & alpha beta-scan \\
    %     \midrule
    %     Stability criterion & Bode plot & Nyquist criterion & Generalized Nyquist criterion & Generalized Nyquist criterion \\
    %     \hline
    %     Parameter & \( Z_{g}^{ps}(f) \) and \( Z_{c}^{ps}(f) \) & \( L(f) \) & \( L(f) \) in dq-frame & \( L(f) \) in alpha beta-frame \\
    %     \hline
    %     Prediction reliability & Good, Depends on impedance model accuracy & Good, Depends on impedance ratio accuracy & Very good, captures dq coupling Depends on eigenvalue accuracy & Depends on eigenvalue accuracy\\
    %     \hline
    %     Stability margin & yes , visualized & yes & yes & yes \\
    %     \hline
    %     Oscillation frequency & Magnitude intersection & intersection with the unit circle & intersection of dominant eigenlocus with the unit circle & intersection of dominant eigenlocus with the unit circle \\
    %     \hline
    %     Computational burden &  Lowest & & Generally, 2xx equivalent ps-scan Here, equal to equivalent ps-scan& 2xx equivalent ps-scan \\
    %     \bottomrule
    %     \end{tabular}
    % \end{table}

    \subsection{Perturbation Type: Voltage or Current}
    The scanning of grid-following Inverter-Based Resources (IBRs) can be conducted using perturbations in either voltage or current. Essentially, the two approaches produce results that are almost identical \cite{jacobsComparativeStudyFrequency2023}. In practical applications, the technique of injecting current is more commonly utilized for frequency scanning on the grid side. In contrast, voltage perturbation is predominantly employed for frequency scanning on the Wind Turbine Generator (WTG) \cite{renRefinedFrequencyScan2016}.
    
\begin{figure*}[t]
    \centering
    % First figure on the left side
    \begin{minipage}{0.49\textwidth}
        \centering
        \includegraphics[width=\textwidth]{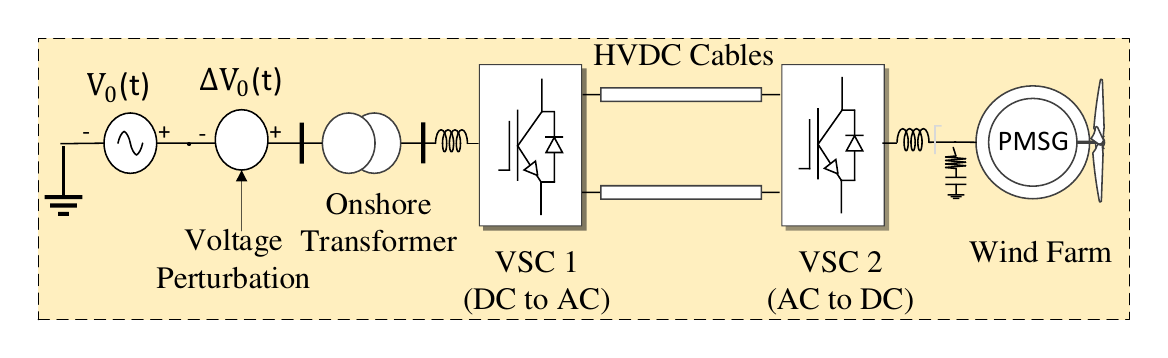}
        \caption{Voltage perturbation}
        \label{fig:vs}
    \end{minipage}
    \hfill
    % Second figure on the right side
    \begin{minipage}{0.48\textwidth}
        \centering
        \includegraphics[width=\textwidth]{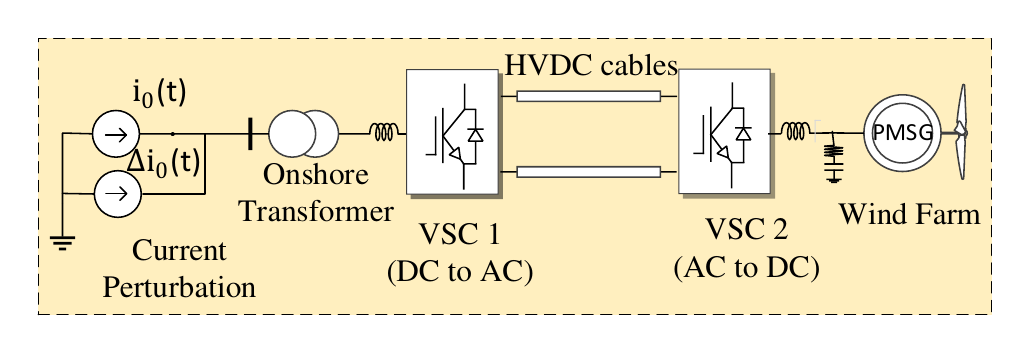}
        \caption{Current perturbation}
        \label{fig:cs}
    \end{minipage}
\end{figure*}

    The scanning mechanism is the same for both cases. However, the connection of voltage and current source is different. It is essential to have the perturb voltage source connected to the series with the grid voltage shown in Fig. \ref{fig:vs}; therefore, the perturbation signal is superposed on the system voltage. On the other hand, current perturbation is injected into the grid in a parallel connection as depicted in Fig. \ref{fig:cs}. A fraction of the perturbation current flows into the grid. If the grid impedance is much smaller than the converter, this might affect the operating point of the system studied \cite{jacobsComparativeStudyFrequency2023}. 
    
    Typically, voltage injection is used for shunt-connected devices like STATCOMs, SVCs, and HVDCs, while current injection is preferred for series-connected devices such as TCSCs and SSSCs. Regardless of the injection method, the scanned subsystem must maintain a stable simulation to capture the frequency response accurately.
     Voltage perturbation scanning is more applicable in systems where voltage stability is a concern, such as weak grids or systems with high renewable energy penetration. In contrast, current perturbation scanning is often used in grids dominated by power electronics or systems with significant inverter-based resources. These two techniques complement each other, often being used together for a thorough assessment of system dynamics and stability.
    
    % Show the schematic of the voltage and current scan. It is essential
    % Also shows the current direction as it is also crucial for scanning.

    \subsection{Perturbation Harmonic}

    Two types of perturbation harmonics are found in the literature, i.e., single-tone and multi-tone. In a single-tone scan, a single-frequency sinusoidal is perturbed and records the output voltage and current. On the other hand, in the multi-tone method, multiple frequency signals are added together to form the perturbing signal. 

    % \subsubsection {\textbf{Single-tone Frequency Scanning}}
    % \label{single-tone-frequency-scanning}
    % In this method, a single-frequency perturb signal is generated in each data record window and is injected into the IBR system. For scanning the IBR, the voltage perturb mechanism is utilized. For scanning the grid, the current perturb method is used. The final impedance scanning results are just the summation of both current and voltage scans. 

    % Since, only single frequency is perturbed in this method, it requires significant amout of time scan for intended range of frequency. For an example, if the intension is to scan for 0--120Hz with a resolution of 0.5 Hz, then 240 perturbation is required in the frequency interval of 0Hz, 0.5Hz, 1.0Hz... 120Hz. 

    % Trevisan et al. \cite{trevisanAnalyticallyValidatedSSCI2021} reference the work of Francis \cite{francisAlgorithmImplementationSystem2011} to discuss the consequences associated with the use of multi-tone injections. However, they also acknowledge the research by Lwin et al. \cite{lwinFrequencyScanConsiderations2019b}, which presents evidence suggesting that simpler, single-tone sinusoidal perturbations are more efficacious when dealing with models that are tailored to specific manufacturers.

    \subsubsection{\textbf{Single-Tone Frequency Scanning}}  
    \label{single-tone-frequency-scanning}

    In single-tone frequency scanning, a single-frequency perturbation signal is injected during each measurement interval. For IBR characterization, a voltage-based perturbation is commonly employed, while current-based perturbation is used for grid characterization. The resulting impedance profile is then obtained by combining the voltage and current scan results.

    A key drawback of this approach is the time required to cover a broad frequency range. For instance, scanning from 0 to 120 Hz at 0.5 Hz intervals would necessitate 240 individual perturbations (at 0 Hz, 0.5 Hz, 1.0 Hz, and so forth). While multi-tone injection can expedite the scanning process, Lwin et al. \cite{lwinFrequencyScanConsiderations2019b} provide evidence that simple single-tone sinusoidal perturbations are often more effective, especially when dealing with proprietary or manufacturer-specific black-box IBR models.
  
    \begin{figure} [!b]
        \centering 
        \includegraphics[width=0.5\textwidth]{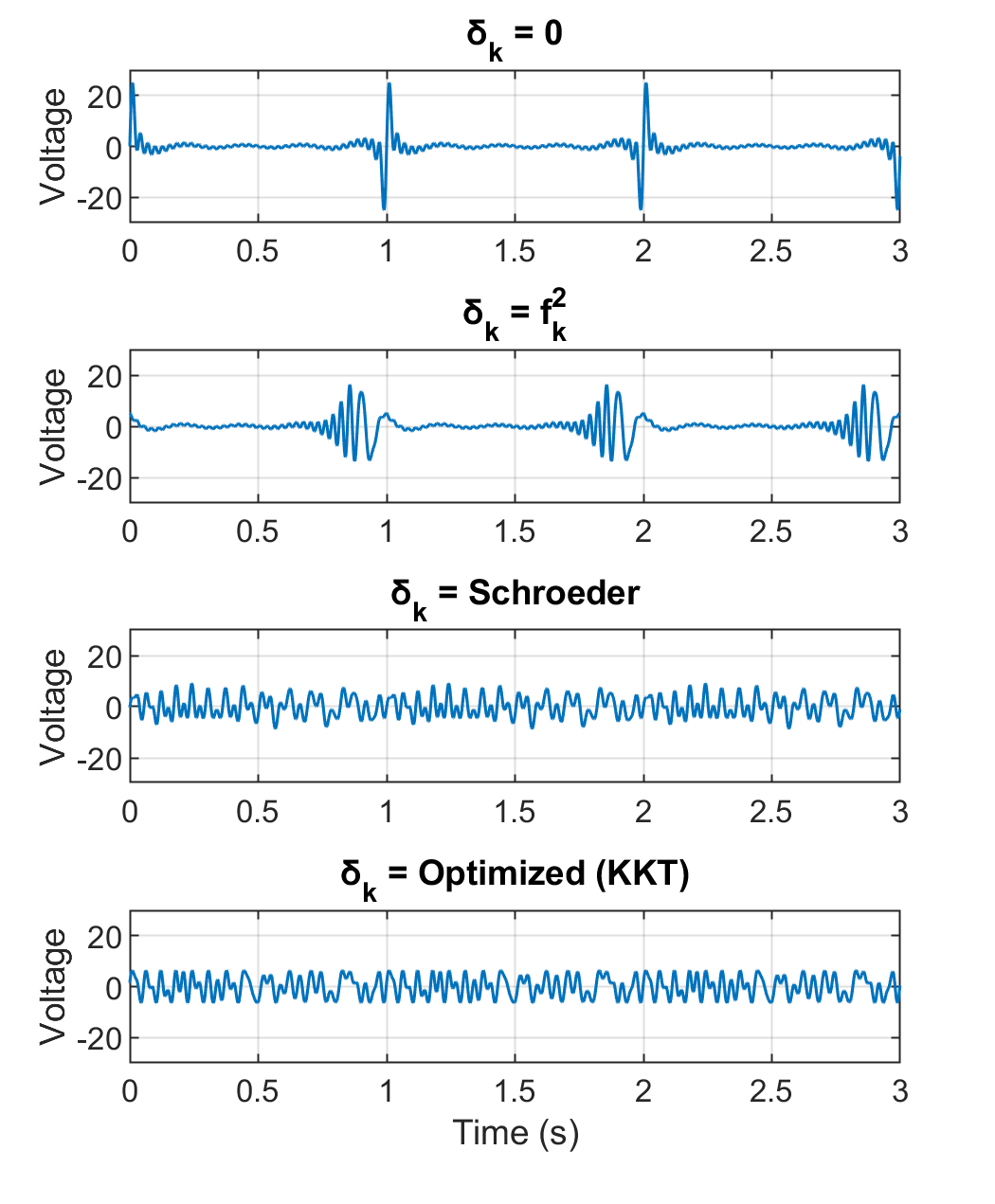} 
        \caption{Multi-tone perturbation signal (Multi-sine Voltage Injection with Different Phase Shifts)} 
        \label{fig:multi-frequency-scan} 
    \end{figure}

\subsubsection{\textbf{Multi-tone Frequency Scan}}
\label{multi-tone-frequency-scan}
Multi-tone frequency scanning is an advanced method used to analyze the behavior of a system across multiple frequencies at the same time. Instead of testing one frequency at a time, as is done in single-tone scanning, this technique involves injecting a signal that contains several different frequencies simultaneously. This allows for a more comprehensive and efficient analysis of the system's frequency response, making it easier to identify potential issues like resonances or instabilities across a wide range of frequencies.
The process begins with generating a multi-tone signal, which is created by combining several sinusoidal waveforms, each representing a different frequency. These frequencies are carefully selected to cover the entire range of interest without interfering with each other. Once the multi-tone signal is ready, it is injected into the system at specific points, such as at a bus or a component that needs to be analyzed. The system's response to this injection is then measured at various locations.
The next step is to transform the measured response from the time domain to the frequency domain using a technique like the Fast Fourier Transform. This step breaks down the complex response into its frequency components, allowing engineers to see how the system behaves at each of the injected frequencies. By examining the amplitude and phase response at these frequencies, they can detect any problematic areas, such as points where the system might become unstable or resonant\cite{shirinzadFrequencyScanBased2021}.
One of the main advantages of multi-tone scanning is its efficiency. Since multiple frequencies are analyzed at once, the scanning process is much faster than single-tone methods. Additionally, this approach provides a broader view of the system's behavior, which is particularly useful in complex systems where problems might occur across a wide frequency range\cite{shirinzadFrequencyScanBased2021}.

% However, multi-tone scanning also comes with challenges. The design of the multi-tone signal must be precise to avoid issues like interference between the tones. Interpreting the results can also be more complex, as the presence of multiple frequencies requires careful analysis to separate and understand the different components of the system's response\cite{shirinzadFrequencyScanBased2021}.

% The researcher \cite{matsuoOptimizedFrequencyScanning2020} presented an optimization technique to improve the crest factor in the muli-frequency scan depicted in the Fig. \ref{fig:multi-frequency-scan}. An optimization algorithm is applied to find the optimum phase so that crest factor can be minimized. Table \ref{table:peak_rms_crest} shows the crest factor, peak and RMS value for 4 different methods. Method 4 shows the minimum peak and crest factor. 
    
However, multi-tone scanning also comes with multiple challenges--primarily accurate signal identification and controlling the crest factor. The design of the multi-tone signal must be precise to avoid interference between tones, and interpreting the results is more complex due to the presence of multiple frequencies, which require careful analysis to separate and understand the system’s response \cite{shirinzadFrequencyScanBased2021}. To address the crest-factor issue, Matsuo \cite{matsuoOptimizedFrequencyScanning2020} presented an optimization technique (Fig. \ref{fig:multi-frequency-scan}) that applies an algorithm to find the optimum phase and minimize the crest factor; Table \ref{table:peak_rms_crest} shows that Method 4 achieves the lowest peak and crest factor.

    \begin{table}[!t]
        \caption{Peak, RMS, and Crest Factor Values}
        \centering
        \begin{tabular}{p{2cm}p{1.6cm}p{1.6cm}p{1.6cm}}
        \hline
        \textbf{Method} & \textbf{Peak} & \textbf{RMS} & \textbf{CF} \\
        \hline
        Method 1 & 25.06 & 3.94 & 6.36 \\
        Method 2 & 16.31 & 3.94 & 4.14 \\
        Method 3 & 9.17  & 3.94 & 2.33 \\
        Method 4 & 6.42  & 3.94 & 1.63 \\
        \hline
        \end{tabular}
        \label{table:peak_rms_crest}
      \end{table}

Finally, Trevisan et al. \cite{trevisanAnalyticallyValidatedSSCI2021} reference the work of Francis \cite{francisAlgorithmImplementationSystem2011} to discuss the consequences associated with the use of multi-tone injections. However, they also acknowledge the research by Lwin et al. \cite{lwinFrequencyScanConsiderations2019b}, which presents evidence suggesting that simpler, single-tone sinusoidal perturbations are more efficacious when dealing with models that are tailored to specific manufacturers.

\subsection{Perturbation Shape}
\label{perturbation_shape}

\begin{figure}[!t]
    \centering

    \begin{minipage}[b]{0.45\textwidth}
        \centering
        \includegraphics[width=\textwidth]{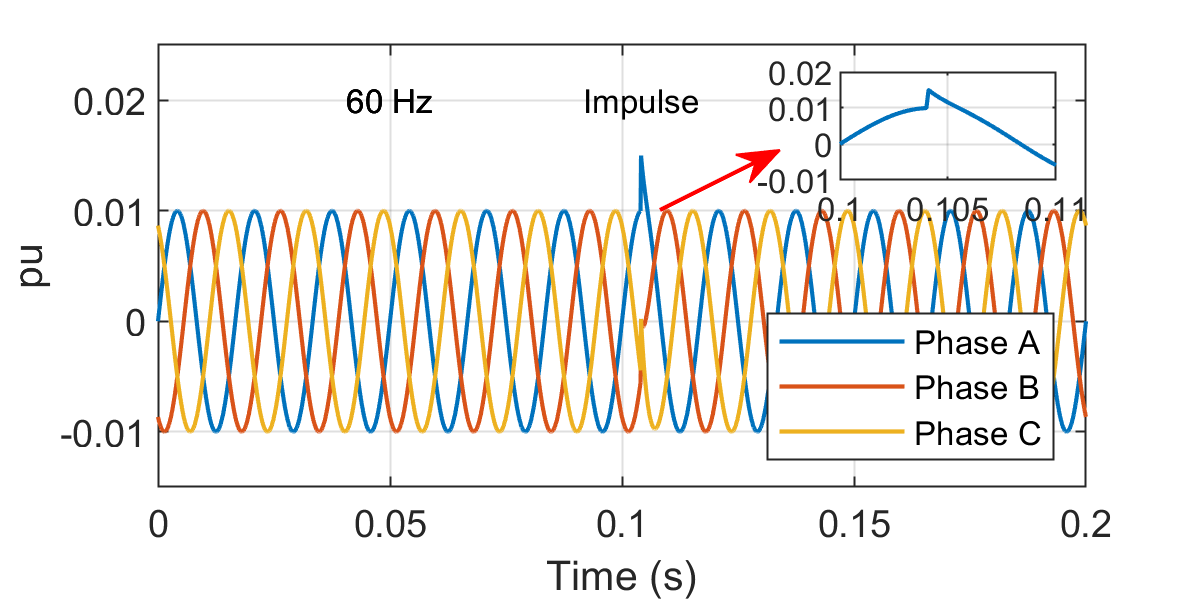}
        \centerline{(a) Impulse perturbation}
        % \label{fig:impulse}
    \end{minipage}

    \begin{minipage}[b]{0.45\textwidth}
        \centering
        \includegraphics[width=\textwidth]{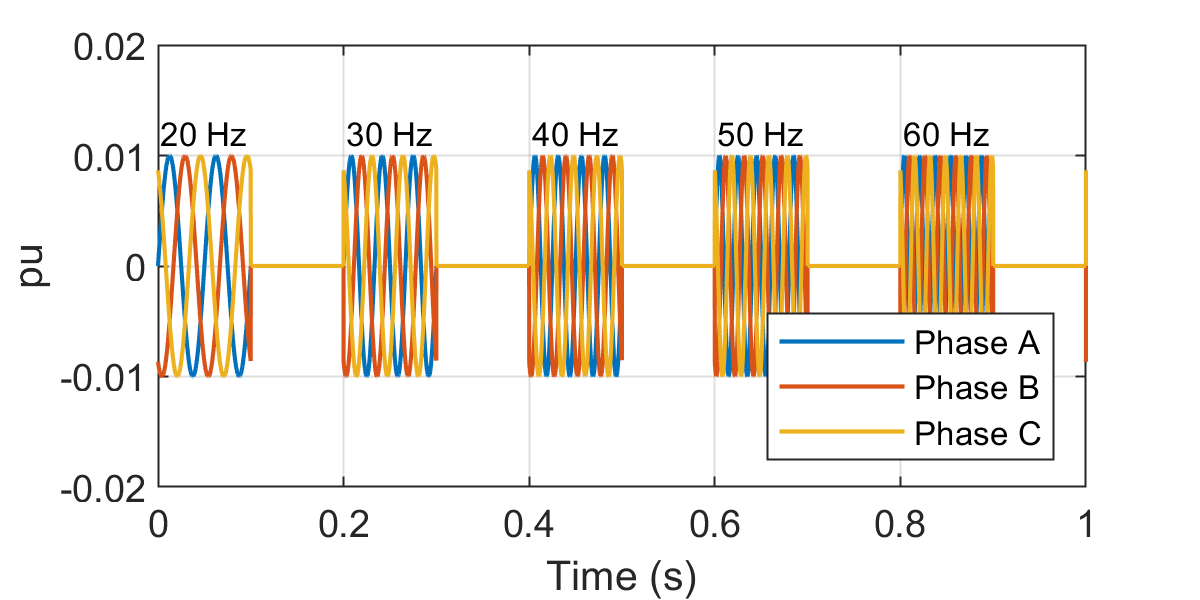}
        \centerline{(b) Step perturbation}
        % \label{fig:step_perturbation}
    \end{minipage}
    % \hfill
    % \vspace{1em}
    \begin{minipage}[b]{0.45\textwidth}
        \centering
        \includegraphics[width=\textwidth]{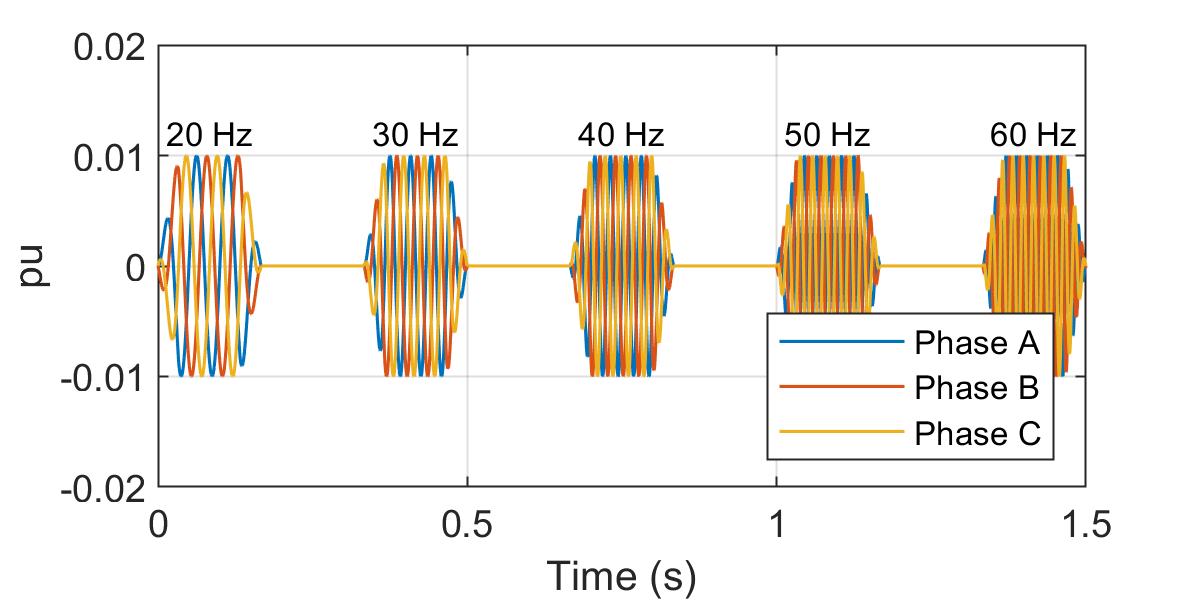}
        \centerline{(c) Ramp perturbation}
        % \label{fig:ramp_perturbation}
    \end{minipage}
    
    % \vspace{1em}
    
    \begin{minipage}[b]{0.45\textwidth}
        \centering
        \includegraphics[width=\textwidth]{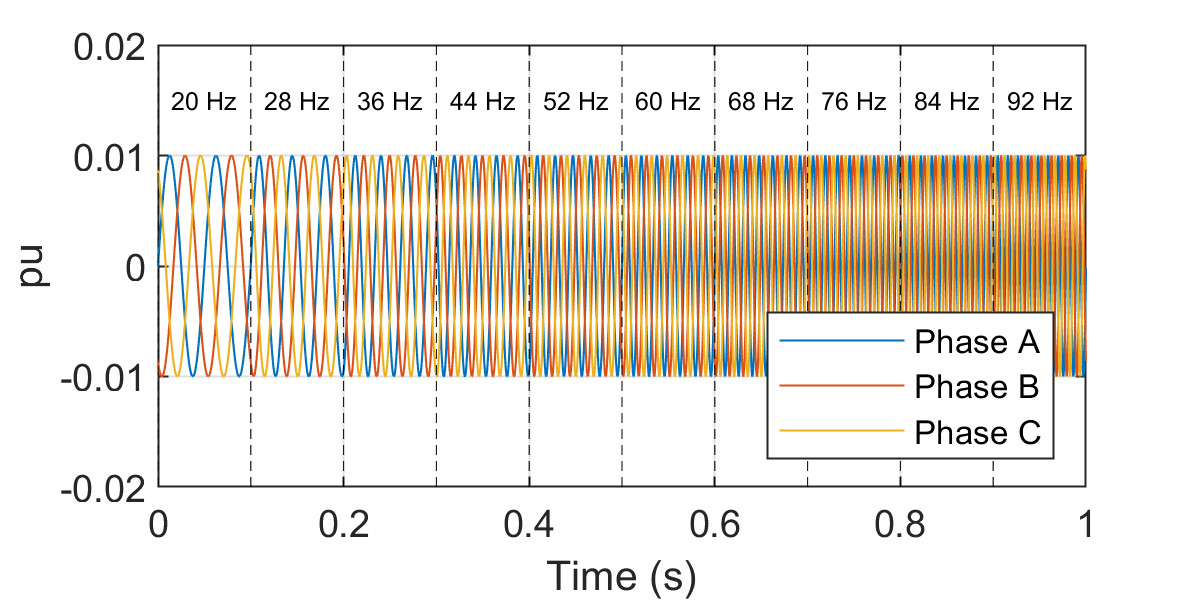}
        \centerline{(d) Chirp perturbation}
        % \label{fig:chirp_perturbation}
    \end{minipage}
    % \hfill
    % \vspace{1em}
    \begin{minipage}[b]{0.45\textwidth}
        \centering
        \includegraphics[width=\textwidth]{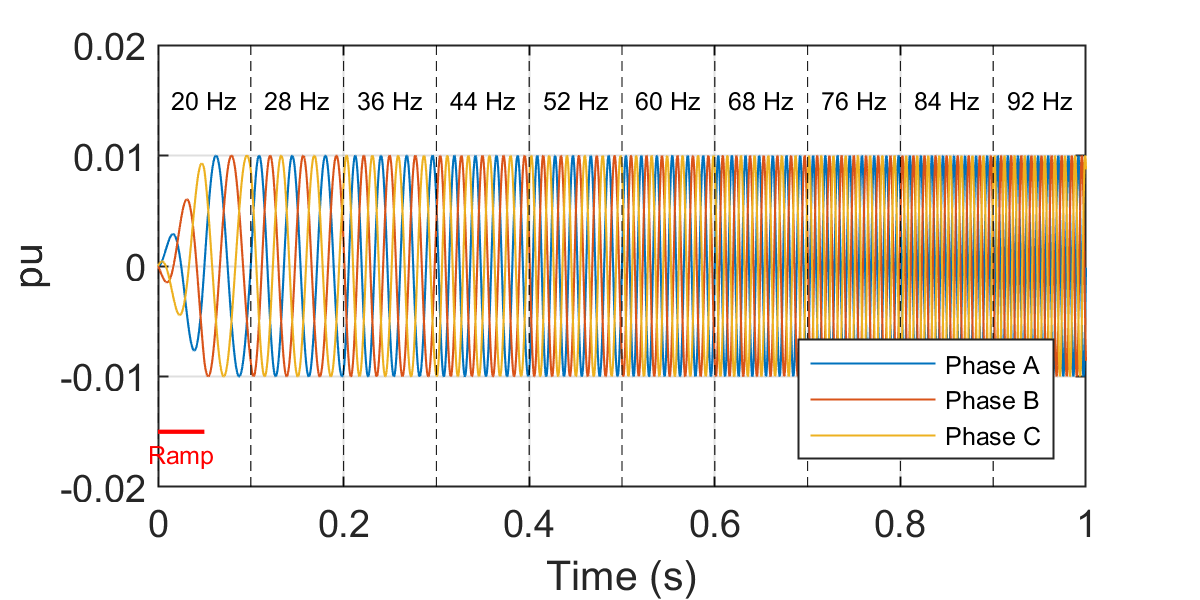}
        \centerline{(e)Ramp+chirp perturbation}
        % \label{fig:ramp_chirp_perturbation}
    \end{minipage}
    % \vspace{1em}

    \caption{Different types of perturbation signals.}
    \label{fig:perturb_shape}
\end{figure}

The perturbation properties, such as its frequency, waveform shape, and magnitude, serve as crucial parameters within the procedure \cite{trevisanAnalyticallyValidatedSSCI2021}. By carefully selecting these parameters, researchers ensure that the induced responses accurately capture the underlying system dynamics, thereby improving the reliability of subsequent parameter estimation, system identification, and model validation processes \cite{aleniusAmplitudeDesignPerturbation2020,martinezPerturbationAnalysisPower2004}. Based on the literature, five primary types of perturbation shapes have been identified: step, ramp, chirp, ramp+chirp, and impulse.

\subsubsection{\textbf{Impulse Perturbation}}
Impulse perturbations, as depicted in Fig.~\ref{fig:perturb_shape}(a), momentarily inject a broadband excitation that encompasses multiple frequencies simultaneously and with equal phase. This sharp, instantaneous signal can quickly reveal certain resonance points and dynamic peculiarities, making it relatively straightforward to implement and interpret in simple systems. However, the large amplitude spikes produced by impulse perturbations may disturb normal system operations and compromise safety. Such abrupt stimuli often generate strong nonlinear effects, thus complicating the extraction of accurate frequency-response information. Given these limitations, the impulsive approach is rarely favored when more nuanced or stable testing methods are available \cite{badrzadehSubsynchronousInteractionWind2012a, evansAssessmentSustainabilityIndicators2009}.

\subsubsection{\textbf{Step Perturbation}}
Step perturbations involve abrupt changes in frequency or amplitude at discrete intervals, creating stable plateau-like segments from which steady-state responses can be easily observed. As illustrated in Fig.~\ref{fig:perturb_shape}(b), this approach simplifies the identification of how a system behaves under different fixed-frequency conditions, making it well-suited for preliminary assessments of system dynamics and stability. Despite these advantages, the sudden transitions associated with step perturbations can activate nonlinear system behaviors and introduce unwanted measurement noise  \cite{zhangResearchSynchronousControl2020}. This complication makes it more challenging to obtain smooth frequency response curves and to distinguish subtle resonance effects. While step perturbations remain a useful starting point, their inherent abruptness often limits their utility in scenarios demanding detailed frequency-response insights \cite{trevisanAnalyticallyValidatedSSCI2021, zhangResearchSynchronousControl2020}.

\subsubsection{\textbf{Ramp Perturbation}}
Ramp perturbations gradually alter the frequency or amplitude of the input signal over time, as shown in Fig.~\ref{fig:perturb_shape}(c). By providing a smoother transition between operating points, ramp signals reduce the likelihood of inciting nonlinearities and result in more stable conditions for testing. This characteristic makes ramp perturbations particularly attractive for systems where preserving stable operation and obtaining subtle resonance information is essential. However, the slower transitions inherent in ramp-based testing can increase the duration of experiments, and the continuous nature of the frequency variation necessitates careful analysis to extract the underlying frequency-response data. Despite these drawbacks, ramp perturbations strike a workable balance between simplicity and stability, rendering them a suitable option for more refined frequency-response analyses \cite{trevisanAnalyticallyValidatedSSCI2021}.

\subsubsection{\textbf{Chirp Perturbation}}
Chirp perturbations, depicted in Fig.~\ref{fig:perturb_shape}(d), continuously vary their frequency content, enabling a comprehensive exploration of a system’s response across a broad frequency spectrum. This approach provides a richer, more continuous mapping of system dynamics than step or ramp perturbations, making it easier to identify resonant frequencies and other frequency-dependent phenomena without resorting to multiple isolated tests. However, the complexity of generating and interpreting chirp signals is higher, and the test duration tends to be longer. Extracting precise frequency-response information requires careful data processing and interpretation, given the continuously changing excitation frequency. While these factors increase the methodological difficulty, chirp perturbations are a powerful method that can yield detailed insights into system behavior that simpler perturbation methods are likely to miss \cite{foyenImpedanceScanningChirps2022, slepskiOptimizationImpedanceMeasurements2009}.

\subsubsection{\textbf{Ramp+Chirp Perturbation}}
The ramp+chirp perturbation, shown in Fig.~\ref{fig:perturb_shape}(e), combines the smoothness of a ramp function with the broad frequency coverage of a chirp signal. By introducing the chirp within a controlled ramp, the method provides a gentler frequency trajectory than a pure chirp, thus reducing abruptness and lowering the risk of triggering nonlinear responses. While this hybrid approach offers improved stability and a more comprehensive examination of the system’s dynamic characteristics, it also elevates complexity in signal design, data processing, and interpretation. Increased testing times and more involved parameter extraction procedures are often necessary. Nevertheless, the ramp+chirp perturbation stands as a sophisticated option for studies requiring both smooth transitions and a wide frequency sweep, delivering high-quality insights into complex system behaviors \cite{trevisanAnalyticallyValidatedSSCI2021}.

\subsection{Explaning Frequency Scanning Result}
    Frequency scanning identifies resonance spots in a system, where the impedance magnitude reaches peaks or troughs. These points are essential because they could indicate operational instability, harmonic amplification, or negative interactions between grid-connected components and their controls. When the reactance crosses from negative to positive, that frequency is called the crossover frequency for series resonance. Similarly, when the reactance crosses from positive to negative, that frequency is called the crossover frequency for parallel resonance \cite{matsuoOptimizedFrequencyScanning2020, renRefinedFrequencyScan2016}. In addition, frequency scan data can illustrate how system strength affects impedance characteristics. For example, impedance profiles in strong grids are relatively flat, and the system behaves stable and well-damped over a wide frequency range meanwhile frequency scanning frequently reveals greater impedance magnitudes, more prominent resonance peaks, and shifts in resonance frequencies for weak grids, suggesting heightened susceptibility to disturbances, emphasizing the possibility for voltage instability and control interactions \cite{nercIntegratingInverterBasedResources2017}. In the case of harmonics, frequency scanning offers information about the propagation and amplification of harmonics within a system. Peaks in impedance profiles at specific harmonic frequencies indicate the presence of harmonic resonance, which can result in power quality concerns, equipment overheating, and operational inefficiencies.

    Existing literature \cite{elfayoumyComprehensiveApproachSubsynchronous2003, nathStudySubSynchronousControl2012, sahniAdvancedScreeningTechniques2012, badrzadehGeneralMethodologyAnalysis2013, chengReactanceScanCrossoverBased2013, guptaFrequencyScanningStudy2013, wenStabilityAnalysisThreephase2014, renRefinedFrequencyScan2016, karaagacSafeOperationDFIGBased2018, liaoGeneralRulesUsing2018, ryggApparentImpedanceAnalysis2017, trevisanAnalyticallyValidatedSSCI2021, fanIdentifyingDQDomainAdmittance2021, lwinFrequencyScanConsiderations2019, liuAnalysisDesignImplementation2020, matsuoOptimizedFrequencyScanning2020, jacobsComparativeStudyFrequency2023, mengNewSequenceDomain2023, ramakrishnaDQAdmittanceExtraction2023, shirinzadFrequencyScanBased} explained frequency results differently. The representation also varies depending on the reference frame the scanning was performed. The interpretation of the results could be categorized mostly by a combined scan- and impedance ratio-based explanation. Both methods have advantages and disadvantages. The combined scan method includes the analysis of Impedance Over Frequency (Z-F), Admittance Over Frequency (Y-F), Phase Angle, and Nyquist Plots. On the other hand, the impedance ratio-based method involves the use of the Generalized Nyquist Criteria and Bode plots for analysis.

    \subsubsection{\textbf{Combined Scans Analysis}}
    
    \begin{figure}[!b]
        \centering 
        \includegraphics[width=0.48\textwidth]{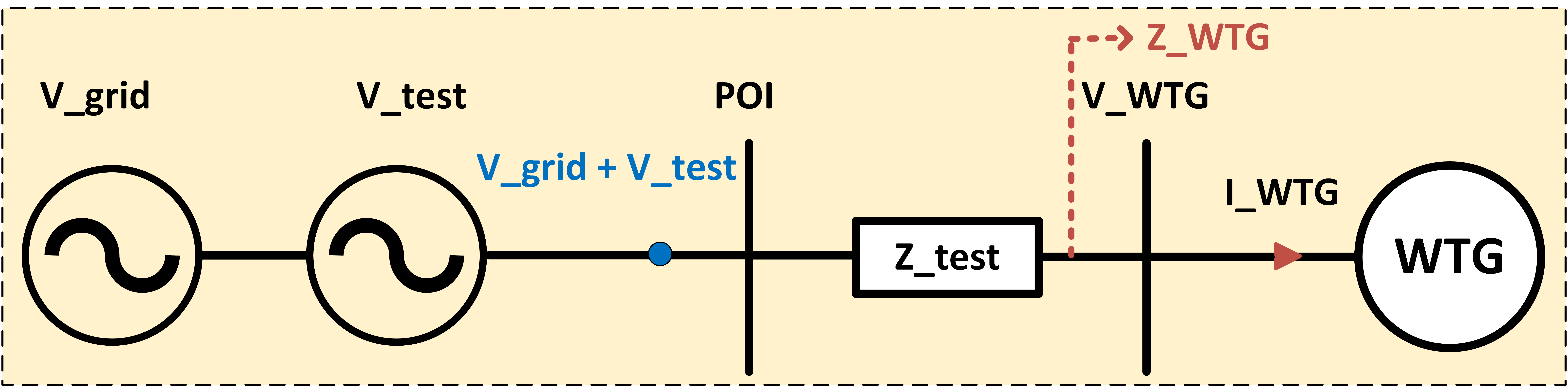} 
        \caption{Combined frequency scan procedure} 
        \label{fig:combined_scan} 
    \end{figure}

    The combined scan method is adopted in many parts of the literature to explain the stability of the overall system \cite{trevisanAnalyticallyValidatedSSCI2021}. Figure \ref{fig:combined_scan} depicts the combined scan procedure. For the point of interconnection (POI), this method computes a frequency-dependent combined impedance of the grid and grid-connected inverters (GCIs). Small voltage or current disturbances are injected into the POI at various frequencies, and the positive sequence frequency-dependent impedances of the grid and the inverter-based resources (IBRs) are extracted from the measured voltages and currents using the Fast Fourier Transform technique. Thus, the total frequency-dependent impedance is determined by adding the two impedances. There are mainly two distinct types of resonance mode: series and parallel resonance. While series resonance transmits the energy within the same branch, parallel resonance energy exchanges between different branches.

    In a series RLC circuit and using $s=j\omega$ the equivalent impedance is:  
    \begin{equation}
      \begin{aligned}
      Z_{\text{eq}_\text{s}} = R + sL + \frac{1}{sC}
      \end{aligned}
      \label{eq:series_imp_laplace}
    \end{equation}
    
    \noindent
    The crossover frequency is where the imaginary part of \( Z_{\text{eq}_\text{s}} \) becomes zero and its derivative w.r.t. \(\omega\) is positive \cite{matsuoOptimizedFrequencyScanning2020}. As \(\omega\) increases, reactance transitions from capacitive to inductive, with resonance when \(\omega L = 1/\omega C\). For an input voltage \(V_{\text{in}}\), the transfer function is:  
   \begin{equation}
      \begin{aligned}
      H_S(s) = \frac{I(s)}{V_{\text{in}}(s)} = \frac{1}{Z_{\text{eq}_\text{s}}} = \frac{Cs}{LCs^2 + RCs + 1}
      \end{aligned}
      \label{eq:series_transferfunc}
    \end{equation}

    \noindent
    The system is unstable if the poles of the transfer function (eqn: \ref{eq:series_transferfunc}) are located in the right-half plane. Since the denominator of (eqn: \ref{eq:series_transferfunc}) is a quadratic equation, the real part of the poles is \( \frac{-R}{2L} \). As a result, the stability of a series circuit would correlate with the resistance value at the resonance frequency.

    % As a result, the series resonance mode would lose the stability if the resistance of the circuit at the crossover frequency is negative.
        
    On the other hand, parallel resonance would be identified if the reactance goes from positive (inductive) to negative (capacitive) when the frequency \( \omega \) grows and it occurs when the inductive and capacitive admittances cancel each other. For parallel resonance, the equivalent impedance \( Z_{\text{eq}_\text{p}} \) comprising a resistor (\( R \)), inductor (\( L \)), and capacitor (\( C \)) is calculated using the reciprocal of the total admittance:

    \begin{equation}
      \begin{aligned}
      \frac{1}{Z_{\text{eq}_\text{p}}} = \frac{1}{R} + \frac{1}{j\omega L} + j\omega C
      \end{aligned}
      \label{eq:parallel_imp}
    \end{equation}
    
    % Where:
    % \begin{align*}
    % Y_R & = \frac{1}{R} \quad \text{(resistive admittance),} \\
    % Y_L & = \frac{1}{j\omega L} \quad \text{(inductive admittance),} \\
    % Y_C & = j\omega C \quad \text{(capacitive admittance).}
    % \end{align*}

    \noindent
    Considering the input current source \( I_{\text{in}} \), the transfer function of the parallel circuit would be:

    \begin{equation}
      \begin{aligned}
      H_P(s) = \frac{V(s)}{I_{\text{in}}(s)} = Z_{\text{eq}_\text{p}} = \frac{RLs}{RLCs^2 + Ls + R}
      \end{aligned}
      \label{eq:parallel_transferfunc}
    \end{equation}

    \noindent
    The real component of the poles of the parallel transfer function (Equation \ref{eq:parallel_transferfunc}) could be calculated as \( \frac{-L}{2RLC} \). If the resistance of the parallel circuit is less than zero at the crossover frequency, the poles would be on the right-half plane, causing the parallel resonance mode unstable. Consequently, combined scan analysis can signify system stability depending on the polarity of the resistance value at the resonance frequency. 

    \begin{figure}[!t]
    \centering
    \begin{minipage}{0.48\textwidth}
        \centering

        \resizebox{0.85\textwidth}{!}{%
        \begin{circuitikz}
            \tikzstyle{every node}=[font=\LARGE]
            
            % Vertical connections on the top
            \draw [short] (2.25,10.75) -- (2.25,9.75);
            \draw (2.25,10.25) to[L=$L_1$, l_=$20\,\text{mH}$] (4.75,10.25);
            \draw (4.75,10.25) to[R=$R_1$, l_=$0.05\,\Omega$] (6.75,10.25);
            
            % Vertical connections on the bottom
            \draw [short] (2.25,8.5) -- (2.25,7.5);
            \draw (2.25,8) to[L=$L_2$, l_=$10\,\text{mH}$] (4.6,8);
            \draw (4.6,8) to[C=$C_1$, l_=$3.5\,\text{mF}$] (5.7,8);
            \draw (5.7,8) to[R=$R_2$, l_=$0.01\,\Omega$] (7.75,8);
            
            % Horizontal connections between top and bottom
            \draw (6.75,10.25) to[short] (7.75,10.25);
            \draw (7.75,10.25) to[short] (7.75,8);
            \draw (7.75,8) to[short] (9.5,8);
            
            % Components in the second dashed rectangle
            \draw (9.5,8) to[L=$L_3$, l_=$1\,\text{mH}$] (11.25,8);
            \draw (11.25,8) to[R=$R_3$, l_=$-1\,\Omega$] (13,8);
            \draw [short] (13,8.5) -- (13,7.5);
            
            % Dashed rectangles for grouping
            \draw [dashed] (1.9,10.9) rectangle (8.25, 6.8);
            \node at (5,6) {sub1};  % Add label under first rectanglefirst rectangle
            \draw [dashed] (8.75,10.9) rectangle (13.36,6.8);
            \node at (11,6) {sub2};  % Add label under second rectangle
        \end{circuitikz}
        }%
           
        % \caption{Placeholder}
        \centerline{(a)}
       
    \end{minipage}\hfill
    \begin{minipage}{0.48\textwidth}
        \centering
        \includegraphics[width=1.0\textwidth]{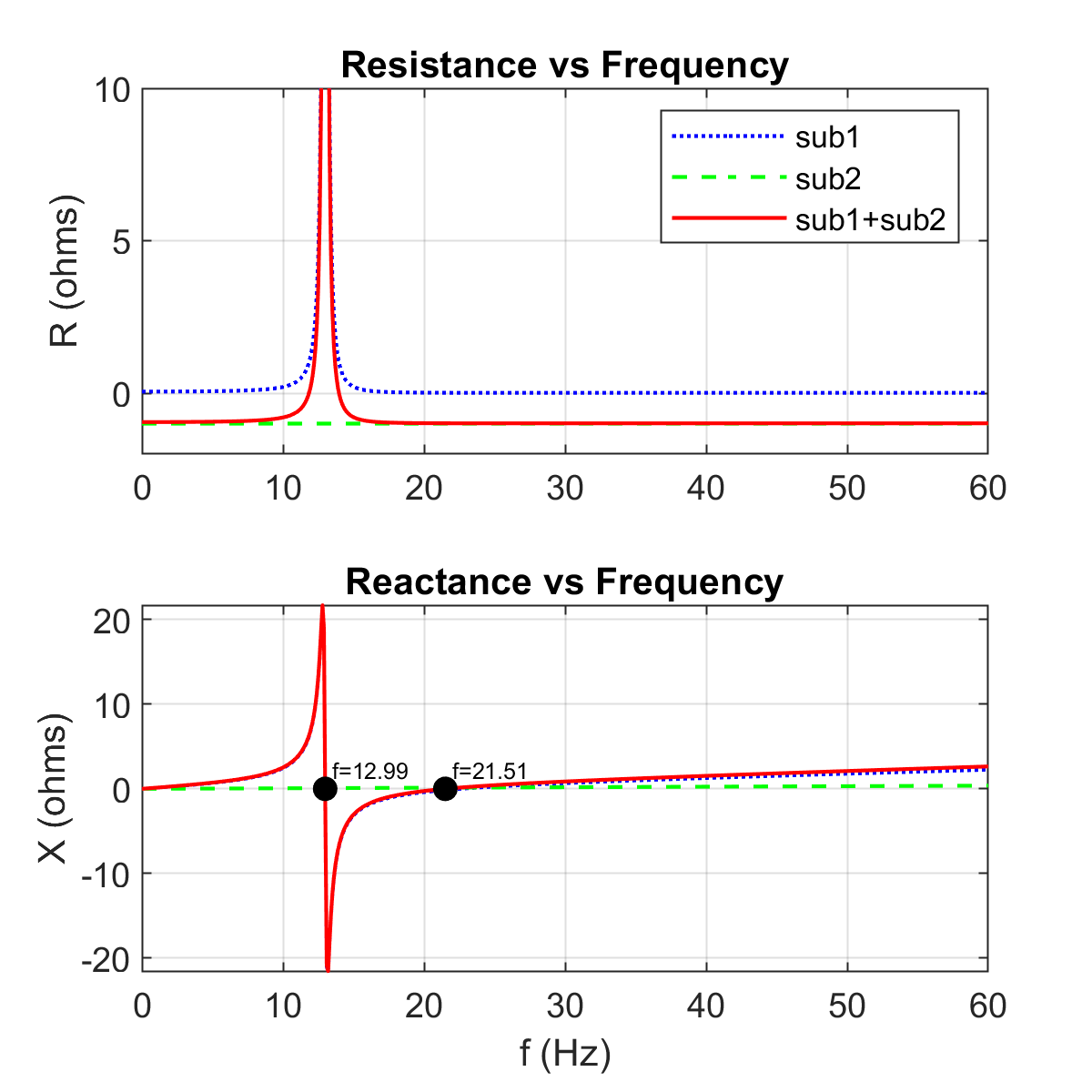}
        % \caption{Figure2Caption}
        \centerline{(b)}
        % \label{fig:figure2}
    \end{minipage}

    \caption{Combined scan result explanation mechanism (a) a simple circuit to demonstrate resonance (b) series and parallel resonance} 
    \label{fig:fscan_explain_ckt}
\end{figure}
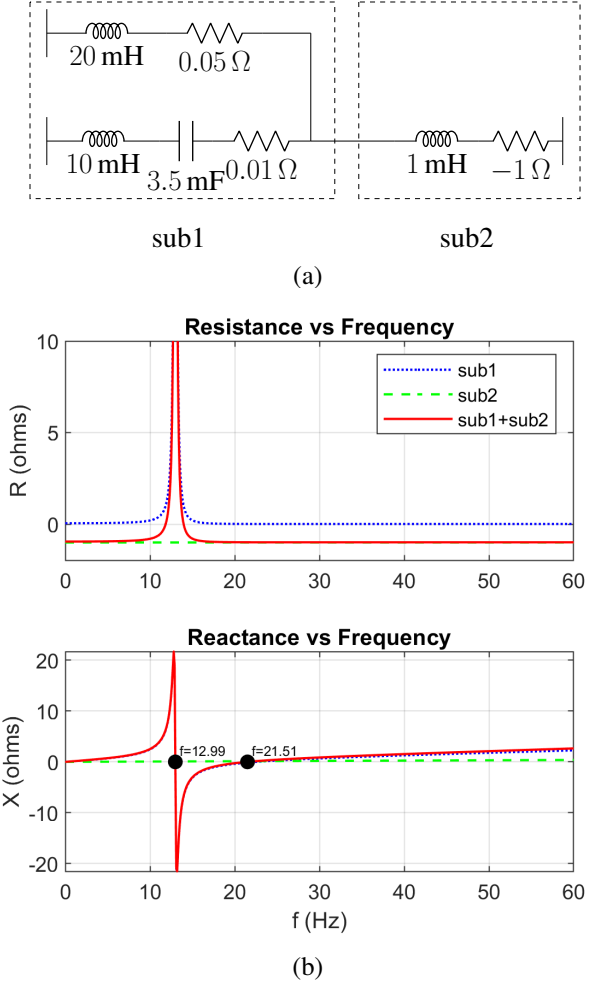

    To understand combined scan result, ref. \cite{renRefinedFrequencyScan2016} explained using a simple circuit demonstrated in Fig. \ref{fig:fscan_explain_ckt} (a). These circuit parameters are hypothetically chosen to study the combined scan analysis. The example circuit operating with 60 Hz fundamental frequency comprises two subsystems. Fig. \ref{fig:fscan_explain_ckt} (b) presents the reactance and resistance of the circuit with respect to frequency. Then the total impedance is the summation of the two and can be evaluated to gain insight to stability. Evaluating the impedance of an electrical system can help identify unstable resonance modes. For example, the circuit in Fig. \ref{fig:fscan_explain_ckt} has two resonance modes at 12.99 Hz and 21.51 Hz where the reactance X crosses zero. The one at 12.99 Hz is a shunt resonance mode as X crosses zero from positive to negative, while the one at 21.51 Hz is a series resonance mode as X crosses zero from negative to positive. Evaluating the series resonance mode shows that it is unstable as the corresponding resistance R is negative. This can be understood by approximating the circuit as an equivalent RLC series branch. Considering the characteristic equation of such branch, negative resistance means that the pole is in the right half plane, therefore the circuit is unstable. 
    
    % While the combined scans analysis technique would be applied to identify multiple resonance modes and interaction effects, the purpose of the impedance ratio-based stability analysis in the following section is to evaluate the stability margin of the system based on the impedance ratio of the source impedance and the load impedance.

    \subsubsection{\textbf{Impedance Ratio Based Stability Analysis}}

\begin{figure*}[htbp]
        \centering
        % First subplot
        \begin{minipage}[b]{0.32\textwidth}
            \centering
            \includegraphics[width=0.95\textwidth]{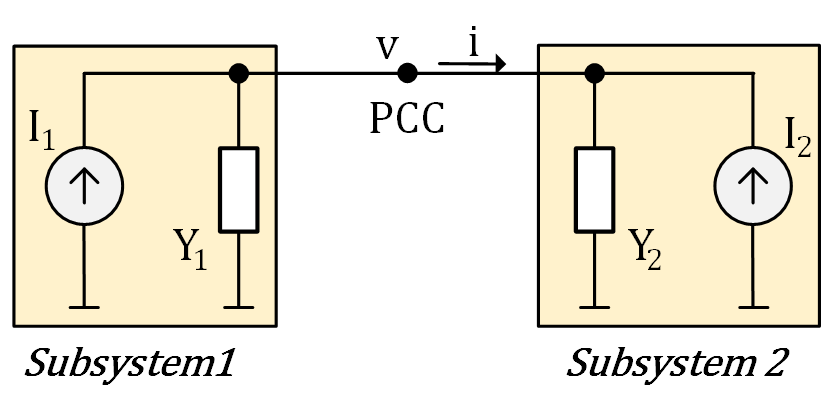}
            \caption*{(a) Norton circuit representation}
            % Your first plot content here
        \end{minipage}
        % \hfill  % Add horizontal spacing between minipages
        % Second subplot
        \begin{minipage}[b]{0.32\textwidth}
            \centering
            \includegraphics[width=0.95\textwidth]{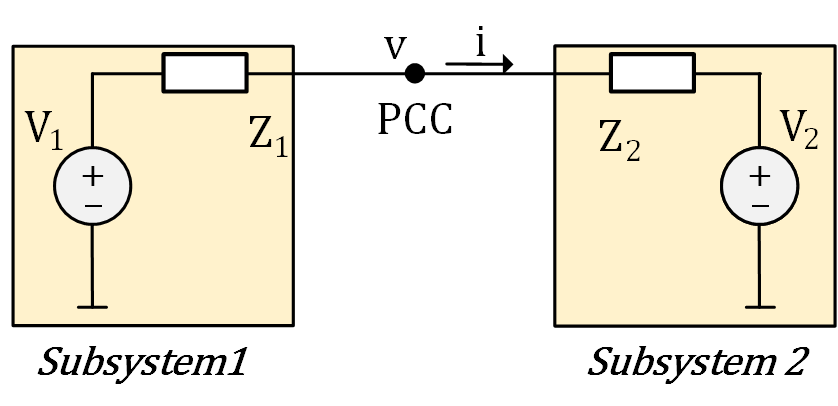}
            \caption*{(b) Thevenin circuit representation}
            % Your second plot content here
        \end{minipage}
        % \hfill  % Add horizontal spacing between minipages
        % Third subplot
        \begin{minipage}[b]{0.32\textwidth}
            \centering
            \includegraphics[width=0.95\textwidth]{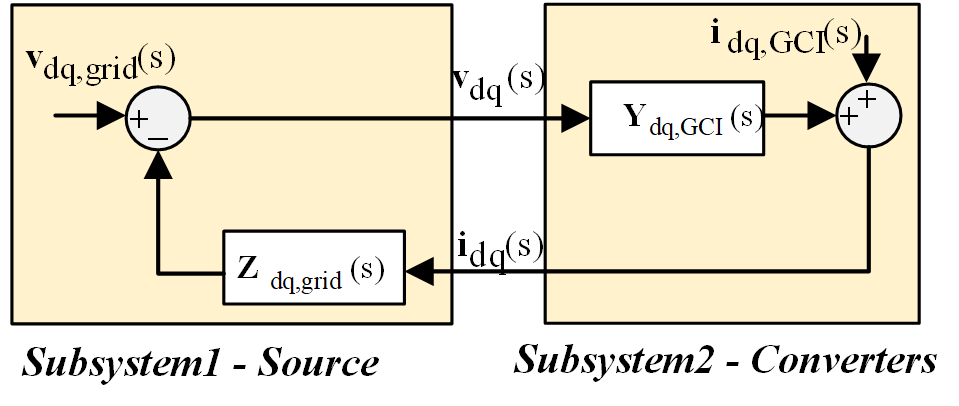}
            \caption*{(c) dq-ISBA circuit representation}
            % Your third plot content here
        \end{minipage}
        % \caption{Cascaded system equivalent circuit (a) Norton circuit (b) Thevenin circuit (c) Block diagram equivalent of the dq-IBSA circuit representation}
        \caption{Cascaded system equivalent circuit}
        \label{fig:imp_ratio_equivalent}
    \end{figure*}

    In power system analyses, it is often practical to partition the network into two subsystems: the element under investigation and the rest of the grid. By representing each subsystem as an equivalent impedance or admittance, an impedance ratio between them can be established \cite{sunImpedanceBasedStabilityCriterion2011,mateu-barriendosOscillatoryFrequencyCharacterization2023}. This impedance ratio has been demonstrated to correspond to the open-loop gain of the system \cite {liaoGeneralRulesUsing2018, yoonHarmonicStabilityAssessment2016}. Each subsystem can be equivalent to the Thevenin or Norton circuit, as shown in the single-line circuit examples of Fig. \ref{fig:imp_ratio_equivalent} (a)--(b). The voltage at the PCC can be calculated as for the Norton circuit:
    \begin{equation}
      \begin{aligned}
      V & =I_1\frac{1}{Y_1+Y_2} +I_2\frac{1}{Y_1+Y_2} =I_1 \frac{Z_1}{1 + Z_1 Y_2} + I_2 \frac{Z_1}{1 + Z_1 Y_2}  \\
      & = I_1 \frac{Z_2}{1 + Z_2 Y_1} + I_2 \frac{Z_2}{1 + Z_2 Y_1}
      \end{aligned}
      \label{eq:thevenin_loop_gain}
    \end{equation}

    % \begin{figure}[!t] 
    %     \centering 
    %     \includegraphics[width=0.48\textwidth]{Figures/Minh_Manh_Figures/Figure 16. Impedance-based representation of cascade systems-Minh.png} 
    %     \caption{Cascaded system equivalent circuit (a) Norton circuit (b) Thevenin circuit} 
    %     \label{fig:thevenin_norton_gci_grid_ckt} 
    % \end{figure}
    
    \noindent
    Similarly, the voltage at the PCC for the Thevenin circuit can be calculated: 
    \begin{equation}
      \begin{aligned}
      V & =\frac{Z_2 V_1 + Z_1 V_2}{Z_1 + Z_2} = \frac{V_1 + \frac{Z_1}{Z_2} V_2}{1 + \frac{Z_1}{Z_2}} = \frac{\frac{Z_2}{Z_1} V_1 + V_2}{1 + \frac{Z_2}{Z_1}}
      \end{aligned}
      \label{eq:norton_loop_gain}
    \end{equation}
    
    \noindent
   In both Thevenin and Norton representations, the impedances of the subsystems are equivalent, such that \( Z_2 = Y_2^{-1} \) and \( Z_1 = Y_1^{-1} \), where \( Y_1 \) and \( Y_2 \) are the admittances of the respective subsystem. The impedance ratio \( Z_1 Y_2 \) serves as the system's open-loop transfer function and is utilized for stability analysis \cite{wangModelingAnalysisHarmonic2014}.
   
   System stability can be evaluated by applying the Nyquist Stability Criterion (NSC) to \( Z_1 Y_2(s) \). The assumption that \( Z_1 Y_2(s) \) has no right-half plane (RHP) poles allows for stability analysis using only the Nyquist plot. Under these conditions, the system is stable if and only if the Nyquist plot of the impedance ratio does not encircle the critical point $(-1, j 0)$.

    The Impedance-Based Stability Criterion (IBSC) plays a pivotal role in analyzing the interactions between power converters and passive components \cite{zhangImpedanceBasedLocalStability2015, jiansunSmallSignalMethodsAC2009}. The IBSC uses terminal impedance/admittance characteristics of the grid’s impedance \( Z_{\text{grid}} \) and the converters admittance \( Y_{\text{CGI}} \) to find out the interconnected system stability. Referring to Fig. \ref{fig:imp_ratio_equivalent}(a)--(b), Subsystem 1 is considered as the Source, and Subsystem 2 as Converters. A graphical representation of this analysis is depicted in Fig. \ref{fig:imp_ratio_equivalent}(c).The stability of the interconnected system can be evaluated using the minor loop gain, defined matrix terms as:
    
    \begin{equation}
        \boldsymbol{L}(s) = \boldsymbol{Z}_{dq, \text{grid}}(s) \boldsymbol{Y}_{dq, \text{GCI}}(s) = \frac{\boldsymbol{Y}_{dq, \text{GCI}}(s)}{\boldsymbol{Y}_{dq, \text{Grid}}(s)}
        \label{eq: grid-CGI-L(s)-gain}
    \end{equation}
    
    \noindent
    
    %$L = Z_{\text{grid}} \cdot Y_{\text{CGI}} = \frac{Y_{\text{CGI}}}{Y_{\text{Grid}}}$
    
    % \begin{figure}[!ht]
    %     \centering
    %     \begin{minipage}{0.35\textwidth}
    %         \centering
    %         \includegraphics[width=1.0\textwidth]{dq-IBSA circuit.png}
    %         \caption{Block diagram equivalent of the dq-IBSA circuit representation}
    %         \label{fig:grid_gci_loop_gain}
    %     \end{minipage}
    % \end{figure}

    %  \begin{equation}
    %   \begin{aligned}
    %   L = Z_{\text{grid}} \cdot Y_{\text{CGI}} = \frac{Y_{\text{CGI}}}{Y_{\text{Grid}}}
    %   \end{aligned}
    %   \label{eq: grid-CGI-loop-gains}
    % \end{equation}
    
    % \noindent
    Within stability analyses, both the magnitude criterion \( |Y_{\text{CGI}}| > |Y_{\text{Grid}}| \) and the negative phase angle crossover criterion (\( \angle Y_{\text{CGI}} - \angle Y_{\text{Grid}} = -\pi \pm 2\pi N \)) must be satisfied \cite{liaoGeneralRulesUsing2018, yoonHarmonicStabilityAssessment2016}. To apply this method effectively, certain prerequisites are required. The converter should maintain stability when it is unloaded or short-circuited to ground in the case of a current-controlled converter—and the grid must remain stable when the converter is disconnected \cite{bakhshizadehImprovingImpedanceBasedStability2018}. Additionally, a significant challenge arises if the network contains nonlinear elements. In such scenarios, the impedance scan must be performed at the same operating point as when the converter is connected to the network, which can be difficult to achieve after disconnecting the converter \cite{shahReversedImpedancebasedStability2023}.

    Control theory emphasizes the importance of open-loop gain in stability analysis, especially when determining phase and gain margins. A system may become unstable if the impedance ratio introduces a phase shift of -180 degree at a frequency where the gain magnitude is equal to or exceeds unity. To extend the Nyquist stability criterion to Multiple Input Multiple Output (MIMO) systems, Gershgorin bands are utilized. These bands help identify critical gains that govern the gain and phase margins of closed-loop systems \cite{wengkhuenhoDirectNyquistArray2000, turnerCaseStudyApplication2013}. The dq-scanning technique, integral to MIMO systems, aids in stability prediction through the Impedance-Based Stability Assessment (IBSA) theory. This method generates a 2x2 admittance matrix for the converter and a 2x2 impedance matrix for the grid at each evaluated frequency \cite{ryggModifiedSequenceDomainImpedance2016}.    A transfer function matrix in the dq-domain, as described in Equation \ref{eq: grid-CGI-L(s)-gain}, is represented using the following notation:

    % \begin{equation}
    %   \boldsymbol{L}(s) = \boldsymbol{Z}_{dq, \text{grid}}(s) \boldsymbol{Y}_{dq, GCI}(s)
    %   \label{eq: grid-CGI-L(s)-gain}
    % \end{equation}

    % \noindent
    % A transfer function matrix in dq is represented by the following notation:

    \begin{equation}
        \begin{aligned}
            \boldsymbol{L}(s) = 
            \begin{bmatrix}
                L_{dd}(s) & -L_{qd}(s) \\
                L_{dq}(s) & L_{qq}(s)
            \end{bmatrix}
        \end{aligned}
        \label{eq: grid-CGI-matrix}
    \end{equation}

     \noindent
    where the negative sign in the (1,2) element is chosen to be consistent with the definition of complex vectors. When implementing the generalized Nyquist theorem to expand from SISO to MIMO systems, it takes into account multiple complex contours. These contours show the eigenvalue \( \lambda_i(s) \) of \( \boldsymbol{L}(s) \), evaluated at \( s = j\omega \). The Nyquist paths are then formed by the eigen-loci of:
    
    \begin{equation}
      \begin{aligned}
      0 = \det(\lambda I_N - \boldsymbol{L}(s))
      \end{aligned}
      \label{eq: grid-CGI-matrix-det}
    \end{equation}

\begin{figure}[!ht]
        \begin{minipage}{0.48\textwidth}
            \centering
            \includegraphics[width=1.0\textwidth]{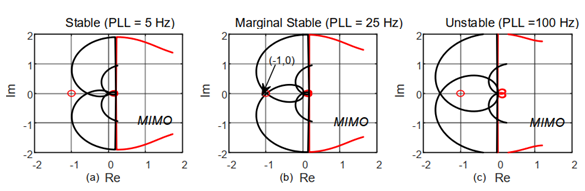}
            \caption{Example: Stability Assessment using MIMO Models \cite{aminNyquistStabilityCriterion2019}}
            \label{Nyquist_example_MIMO}
        \end{minipage}
    \end{figure}

    Equation \ref{eq: grid-CGI-matrix-det} defines the open-loop transfer function matrix \( \boldsymbol{L}(s) \) in an electrical system, which includes both impedance and admittance matrices. To assess how impedance and admittance individually influence closed-loop stability, it is essential to separate the two eigenvalues \( \lambda_1,_2(s) \) as functions of their respective impedance and admittance components. By substituting \ref{eq: grid-CGI-L(s)-gain} into \ref{eq: grid-CGI-matrix-det}, where \( \boldsymbol{L}(s) \) is a \( 2 \times 2 \) transfer function matrix, separating the eigenvalues requires that one of the transfer function matrices (either \( \boldsymbol{Y}_{dq, \text{GCI}}(s) \) or \( \boldsymbol{Z}_{dq, \text{grid}}(s) \)) must have zero entries for both \( L_{dd}(s) \) and \( L_{qq}(s) \) or both \( L_{dq}(s) \) and \( L_{qd}(s) \) (diagonal or antidiagonal). The Nyquist plot of the return ratio then gives useful information on the system's stability, as shown in Figure \ref{Nyquist_example_MIMO}.

    Despite its utility, GNC has limitations, primarily providing only qualitative stability results (stable or unstable) without detailed explanations of instability causes. Furthermore, getting all of the required impedance responses is difficult due to the computing effort involved and the need to individually scan the network impedance with IBRs. To overcome these limitations, alternative stability criteria have been proposed. To support power networks with high levels of IBRs, the National Renewable Energy Laboratory (NREL) developed a Reversed Impedance-Based Stability Criterion (RIBSC) and the Grid Impedance Scan Tool (GIST) \cite{shahReversedImpedancebasedStability2023a}. The Vector Fitting (VF) approach was also used to improve the IBSC \cite{bakhshizadehImprovingImpedanceBasedStability2018}. Moreover, stability criteria based on the impedance-frequency characteristics of the impedance matrix determinant \cite{liuSubsynchronousInteractionDirectDrive2017c}, and the nodal admittance matrix (NAM)-based criterion—which preserves the system's inherent structure by leveraging the system admittance matrix—have been presented \cite{qiaoSmallSignalStabilityAnalysis2024, liStabilityAnalysisLocation2021}. These approaches offer more comprehensive insights into system stability.

\section{Discussion}
\label{discussion} 
 This literature has explored various methodologies for analyzing SSO and system stability in power networks, with a focus on frequency scanning techniques and perturbation-based methods. Frequency scanning methods, categorized into positive-sequence, dq-frame, and $\alpha \beta$-frame scans, emerged as critical tools for understanding the dynamic interactions between grid components and IBRs. While the positive-sequence scan provides quick insights-offering an overall, high-level view of the system's resonance points and potential areas of instability without delving into detailed dynamics-it is less accurate for converter-based systems, as it simplifies the system into a single sequence. The dq-frame scan is more precise for such systems, as it captures the coupling between axes. Multi-tone frequency scanning, though efficient for wide-range frequency analysis, presents challenges in designing perturbation signals and interpreting results due to interference between frequencies. The discussion around perturbation shapes, including ramp, chirp, and impulse perturbations, further emphasizes the importance of selecting appropriate perturbations to prevent non-linear responses and ensure accurate system analysis. The analysis of frequency scanning results through combined scan methods and impedance ratio-based approaches offers a robust understanding of the conditions leading to system instability. The review showed that while resonance frequencies indicate critical points in the system, it is the behavior of resistance at these frequencies that ultimately determines whether the system is stable. The combined scan method, which adds grid and IBR impedances, allows for a comprehensive view of system dynamics, whereas the impedance ratio method provides a more control-theory-driven approach using Nyquist Stability Criteria.

\section{Conclusion}
\label{conclusion}
% This paper contributes to a deeper understanding of SSO and the stability challenges posed by the high penetration of IBRs in modern power systems. The paper categorizes stability approaches, discussing the strengths and limitations of white-box, black-box, and gray-box methods while focusing on frequency scanning techniques as crucial tools for diagnosing and mitigating SSOs. The findings underscore the importance of selecting appropriate perturbation shapes and scanning methods, particularly in IBRs, to accurately assess and enhance grid stability. As grids increasingly integrate renewable energy, a combination of advanced analysis methods will be essential to prevent instability and ensure resilient operation.

In conclusion, as power systems rapidly transition toward inverter-based resources, ensuring reliable and stable operation demands increasingly sophisticated  analytical tools and methodologies. This survey has traced the evolution of stability challenges particularly subsynchronous resonance (SSR) and oscillation phenomena by reviewing past events and categorizing widely used modeling approaches into white-box, black-box, and gray-box frameworks. Among these, frequency scanning methods stand out as versatile and powerful tools, offering insights through positive-sequence, dq-frame, and \(\alpha\beta\)-frame analyses, along with various perturbation strategies such as single-tone, multi-tone, step, ramp, chirp, and combined ramp-chirp. When combined with impedance ratio-based interpretations and stability criteria (including Nyquist and Generalized Nyquist methods), frequency scanning enables a detailed understanding of system behavior and potential instability points. Ultimately, selecting appropriate scanning methods, perturbation shapes, and interpretation techniques is critical for operators and researchers seeking to identify, mitigate, and prevent adverse interactions in systems with high levels of renewable penetration.

\bibliographystyle{IEEEtran}
% \bibliography{References}
\bibliography{ref_phd_thesis}

\begin{thebibliography}{100}
\providecommand{\url}[1]{#1}
\csname url@samestyle\endcsname
\providecommand{\newblock}{\relax}
\providecommand{\bibinfo}[2]{#2}
\providecommand{\BIBentrySTDinterwordspacing}{\spaceskip=0pt\relax}
\providecommand{\BIBentryALTinterwordstretchfactor}{4}
\providecommand{\BIBentryALTinterwordspacing}{\spaceskip=\fontdimen2\font plus
\BIBentryALTinterwordstretchfactor\fontdimen3\font minus
  \fontdimen4\font\relax}
\providecommand{\BIBforeignlanguage}[2]{{%
\expandafter\ifx\csname l@#1\endcsname\relax
\typeout{** WARNING: IEEEtran.bst: No hyphenation pattern has been}%
\typeout{** loaded for the language `#1'. Using the pattern for}%
\typeout{** the default language instead.}%
\else
\language=\csname l@#1\endcsname
\fi
#2}}
\providecommand{\BIBdecl}{\relax}
\BIBdecl

\bibitem{ieaTrackingCleanEnergy2022}
IEA, ``Tracking {{Clean Energy Progress}},''
  https://www.iea.org/reports/renewables, 2022.

\bibitem{blancoEconomicsWindEnergy2009}
M.~I. Blanco, ``The economics of wind energy,'' \emph{Renewable and Sustainable
  Energy Reviews}, vol.~13, no.~6, pp. 1372--1382, Aug. 2009.

\bibitem{mostafaeipourRenewableEnergyIssues2009}
A.~Mostafaeipour and N.~Mostafaeipour, ``Renewable energy issues and
  electricity production in {{Middle East}} compared with {{Iran}},''
  \emph{Renewable and Sustainable Energy Reviews}, vol.~13, no.~6, pp.
  1641--1645, Aug. 2009.

\bibitem{evansAssessmentSustainabilityIndicators2009}
A.~Evans, V.~Strezov, and T.~J. Evans, ``Assessment of sustainability
  indicators for renewable energy technologies,'' \emph{Renewable and
  Sustainable Energy Reviews}, vol.~13, no.~5, pp. 1082--1088, Jun. 2009.

\bibitem{grossSubSynchronousGridConditions2010}
L.~Gross, ``Sub-{{Synchronous Grid Conditions}}: {{New Event}}, {{New
  Problem}}, and {{New Solutions}},'' 2010.

\bibitem{ieeeReadersGuideSubsynchronous1992}
IEEE, ``Reader's guide to subsynchronous resonance,'' \emph{IEEE Transactions
  on Power Systems}, vol.~7, no.~1, pp. 150--157, Feb. 1992.

\bibitem{butlerAnalysisSeriesCapacitor1937}
J.~W. Butler and C.~Concordia, ``Analysis of {{Series Capacitor Application
  Problems}},'' \emph{Transactions of the American Institute of Electrical
  Engineers}, vol.~56, no.~8, pp. 975--988, Aug. 1937.

\bibitem{ballanceSubsynchronousResonanceSeries1973}
J.~W. Ballance and S.~Goldberg, ``Subsynchronous {{Resonance}} in {{Series
  Compensated Transmission Lines}},'' \emph{IEEE Transactions on Power
  Apparatus and Systems}, vol. PAS-92, no.~5, pp. 1649--1658, Sep. 1973.

\bibitem{wangReviewEmergingSSR2017}
L.~Wang, X.~Xie, H.~Liu, Y.~Zhan, J.~He, and C.~Wang, ``Review of emerging
  {{SSR}}/{{SSO}} issues and their classifications,'' \emph{The Journal of
  Engineering}, vol. 2017, no.~13, pp. 1666--1670, 2017.

\bibitem{elfayoumyComprehensiveApproachSubsynchronous2003}
M.~Elfayoumy and C.~Moran, ``A comprehensive approach for sub-synchronous
  resonance screening analysis using frequency scanning technique,'' in
  \emph{2003 {{IEEE Bologna Power Tech Conference Proceedings}},},
  vol.~2.\hskip 1em plus 0.5em minus 0.4em\relax Bologna, Italy: IEEE, 2003,
  pp. 626--630.

\bibitem{matsuoOptimizedFrequencyScanning2020}
I.~B.~M. Matsuo, F.~Salehi, L.~Zhao, Y.~Zhou, and W.-J. Lee, ``Optimized
  {{Frequency Scanning}} of {{Nonlinear Devices Applied}} to {{Subsynchronous
  Resonance Screening}},'' \emph{IEEE Transactions on Industry Applications},
  vol.~56, no.~3, pp. 2281--2291, May 2020.

\bibitem{liaoGeneralRulesUsing2018}
Y.~Liao and X.~Wang, \emph{General {{Rules}} of {{Using Bode Plots}} for
  {{Impedance-Based Stability Analysis}}}, Jun. 2018.

\bibitem{trevisanAnalyticallyValidatedSSCI2021}
A.~S. Trevisan, {\^A}.~Mendon{\c c}a, R.~Gagnon, J.~Mahseredjian, and
  M.~Fecteau, ``Analytically {{Validated SSCI Assessment Technique}} for {{Wind
  Parks}} in {{Series Compensated Grids}},'' \emph{IEEE Transactions on Power
  Systems}, vol.~36, no.~1, pp. 39--48, Jan. 2021.

\bibitem{sunImpedanceBasedStabilityCriterion2011}
J.~Sun, ``Impedance-{{Based Stability Criterion}} for {{Grid-Connected
  Inverters}},'' \emph{IEEE Transactions on Power Electronics}, vol.~26,
  no.~11, pp. 3075--3078, Nov. 2011.

\bibitem{badrzadehGeneralMethodologyAnalysis2013}
B.~Badrzadeh, M.~Sahni, Y.~Zhou, D.~Muthumuni, and A.~Gole, ``General
  {{Methodology}} for {{Analysis}} of {{Sub-Synchronous Interaction}} in {{Wind
  Power Plants}},'' \emph{IEEE Transactions on Power Systems}, vol.~28, no.~2,
  pp. 1858--1869, May 2013.

\bibitem{karaagacSafeOperationDFIGBased2018}
U.~Karaagac, J.~Mahseredjian, S.~Jensen, R.~Gagnon, M.~Fecteau, and I.~Kocar,
  ``Safe {{Operation}} of {{DFIG-Based Wind Parks}} in {{Series-Compensated
  Systems}},'' \emph{IEEE Transactions on Power Delivery}, vol.~33, no.~2, pp.
  709--718, Apr. 2018.

\bibitem{buchhagenBorWin1FirstExperiences2015a}
C.~Buchhagen, C.~Rauscher, A.~Menze, and J.~Jung, ``{{BorWin1}} - {{First
  Experiences}} with harmonic interactions in converter dominated grids,'' in
  \emph{International {{ETG Congress}} 2015; {{Die Energiewende}} -
  {{Blueprints}} for the New Energy Age}, Nov. 2015, pp. 1--7.

\bibitem{chengSubsynchronousResonanceAssessment2019}
Y.~Cheng, S.~H.~F. Huang, J.~Rose, V.~A. Pappu, and J.~Conto, ``Subsynchronous
  resonance assessment for a large system with multiple series compensated
  transmission circuits,'' \emph{IET Renewable Power Generation}, vol.~13,
  no.~1, pp. 27--32, 2019.

\bibitem{saadResonancesHarmonicsHVDCMMC2017}
H.~Saad, Y.~Fillion, S.~Deschanvres, Y.~Vernay, and S.~Dennetiere, ``On
  {{Resonances}} and {{Harmonics}} in {{HVDC-MMC Station Connected}} to {{AC
  Grid}},'' \emph{IEEE Transactions on Power Delivery}, vol.~32, no.~3, pp.
  1565--1573, Jun. 2017.

\bibitem{karnikEvaluationCriticalImpact2017}
N.~Karnik, D.~Novosad, H.~K. Nia, M.~Sahni, M.~Ghavami, and H.~Yin, ``An
  evaluation of critical impact factors for {{SSCI}} analysis for wind power
  plants: {{A}} utility perspective,'' in \emph{2017 {{IEEE Power}} \& {{Energy
  Society General Meeting}}}, Jul. 2017, pp. 1--5.

\bibitem{ieeepesWindEnergySystems2020}
I.~PES, ``Wind {{Energy Systems Sub-Synchronous Oscillations}}: {{Events}} and
  {{Modeling}} ({{TR80}}),''
  https://resourcecenter.ieee-pes.org/publications/technical-reports/PES\_TP\_TR80\_AMPS\_WSSO\_070920.html,
  2020.

\bibitem{mulawarmanDetectionUndampedSubSynchronous2011a}
A.~Mulawarman, ``Detection of {{Undamped Sub-Synchronous Oscillations}} of
  {{Wind Generators}} with {{Series Compensated Lines}},'' 2011.

\bibitem{irwinSubsynchronousControlInteractions2011}
G.~D. Irwin, A.~K. Jindal, and A.~L. Isaacs, ``Sub-synchronous control
  interactions between type 3 wind turbines and series compensated {{AC}}
  transmission systems,'' in \emph{2011 {{IEEE Power}} and {{Energy Society
  General Meeting}}}, Jul. 2011, pp. 1--6.

\bibitem{nercReliabilityGuidelineForced2017}
{North American Electric Reliability Corporation (NERC)}, ``Reliability
  guideline: Forced oscillation monitoring and mitigation,'' Sep. 2017,
  accessed: 2023-10-23.

\bibitem{huangVoltageControlChallenges2012a}
S.-H. Huang, J.~Schmall, J.~Conto, J.~Adams, Y.~Zhang, and C.~Carter, ``Voltage
  control challenges on weak grids with high penetration of wind generation:
  {{ERCOT}} experience,'' in \emph{2012 {{IEEE Power}} and {{Energy Society
  General Meeting}}}, Jul. 2012, pp. 1--7.

\bibitem{IEEEPESWindSSO2020}
``{{IEEE PES WindSSO Taskforce}}, ``{{PES TR-80}}: {{Wind Energy Systems
  Subsynchronous Oscillations}}: {{Events}} and {{Modeling}},'' Jul. 2020.

\bibitem{xieCharacteristicAnalysisSubsynchronous2017a}
X.~Xie, X.~Zhang, H.~Liu, H.~Liu, Y.~Li, and C.~Zhang, ``Characteristic
  {{Analysis}} of {{Subsynchronous Resonance}} in {{Practical Wind Farms
  Connected}} to {{Series-Compensated Transmissions}},'' \emph{IEEE
  Transactions on Energy Conversion}, vol.~32, no.~3, pp. 1117--1126, Sep.
  2017.

\bibitem{liuSubsynchronousInteractionDirectDrive2017a}
H.~Liu, X.~Xie, J.~He, T.~Xu, Z.~Yu, C.~Wang, and C.~Zhang, ``Subsynchronous
  {{Interaction}} between {{Direct-Drive PMSG Based Wind Farms}} and {{Weak AC
  Networks}},'' \emph{IEEE Transactions on Power Systems}, vol.~32, no.~6, pp.
  4708--4720, 2017.

\bibitem{liUnstableOperationPhotovoltaic2018}
C.~Li, ``Unstable {{Operation}} of {{Photovoltaic Inverter From Field
  Experiences}},'' \emph{IEEE Transactions on Power Delivery}, vol.~33, no.~2,
  pp. 1013--1015, Apr. 2018.

\bibitem{vietoBehaviorModelingDamping2018}
I.~Vieto, G.~Li, and J.~Sun, ``Behavior, {{Modeling}} and {{Damping}} of a
  {{New Type}} of {{Resonance Involving Type-III Wind Turbines}},'' in
  \emph{2018 {{IEEE}} 19th {{Workshop}} on {{Control}} and {{Modeling}} for
  {{Power Electronics}} ({{COMPEL}})}, Jun. 2018, pp. 1--8.

\bibitem{sunDevelopmentApplicationTypeIII2020}
J.~Sun and I.~Vieto, ``Development and {{Application}} of {{Type-III Turbine
  Impedance Models Including DC Bus Dynamics}},'' \emph{IEEE Open Journal of
  Power Electronics}, vol.~1, pp. 513--528, 2020.

\bibitem{morjariaDeployingUtilityScalePV}
M.~Morjaria, ``Deploying {{Utility-Scale PV Power Plants}} in {{Weak Grids}}.''

\bibitem{shahIdentifyingPotential2024}
\BIBentryALTinterwordspacing
S.~Shah, J.~Lu, and N.~Modi, ``Identifying potential sub-synchronous
  oscillations using impedance scan approach: Preprint.''\hskip 1em plus 0.5em
  minus 0.4em\relax National Renewable Energy Laboratory (NREL), Golden, CO
  (United States), 08 2024. [Online]. Available:
  \url{https://www.osti.gov/biblio/2439741}
\BIBentrySTDinterwordspacing

\bibitem{jalaliSystemStrengthChallenges2021}
A.~Jalali, B.~Badrzadeh, J.~Lu, N.~Modi, and M.~Gordon, ``System strength
  challenges and solutions developed for a remote area of {{Australian}} power
  system with high penetration of inverter-based resources,'' \emph{CIGRE Sci.
  Eng. J.}, pp. 27--37, 2021.

\bibitem{liAssetConditionAnomaly2019}
C.~Li and R.~Reinmuller, ``Asset {{Condition Anomaly Detections}} by {{Using
  Power Quality Data Analytics}},'' in \emph{2019 {{IEEE Power}} \& {{Energy
  Society General Meeting}} ({{PESGM}})}, Aug. 2019, pp. 1--5.

\bibitem{GBPowerSystem}
``{{GB}} power system disruption on 9 {{August}} 2019: {{Final}} report,''
  \emph{Final report}.

\bibitem{AEMOSystemStrength}
``{{AEMO}} {\textbar} {{System}} strength workshop,''
  https://aemo.com.au/en/learn/energy-explained/system-strength-workshop.

\bibitem{wangIdentifyingOscillationsInjected2022}
C.~Wang, C.~Mishra, K.~D. Jones, R.~M. Gardner, and L.~Vanfretti, ``Identifying
  {{Oscillations Injected}} by {{Inverter-Based Solar Energy Sources}},'' in
  \emph{2022 {{IEEE Power}} \& {{Energy Society General Meeting}}
  ({{PESGM}})}.\hskip 1em plus 0.5em minus 0.4em\relax Denver, CO, USA: IEEE,
  Jul. 2022, pp. 1--5.

\bibitem{GPSTESIGWebinar}
``G-{{PST}}/{{ESIG Webinar Series}}: {{Managing Grid Stability}} in a {{High
  IBR Network}} - {{ESIG}},''
  https://www.esig.energy/event/webinar-managing-grid-stability-in-a-high-ibr-network/.

\bibitem{dongAnalysisNovember212023}
S.~Dong, B.~Wang, J.~Tan, C.~J. Kruse, B.~W. Rockwell, and A.~Hoke, ``Analysis
  of {{November}} 21, 2021, {{Kaua}}`i {{Island Power System}} 18-20 {{Hz
  Oscillations}},'' Feb. 2023.

\bibitem{moharanaSSRMitigationWind2012}
A.~Moharana, R.~K. Varma, and R.~Seethapathy, ``{{SSR}} mitigation in wind farm
  connected to series compensated transmission line using {{STATCOM}},'' in
  \emph{2012 {{IEEE Power Electronics}} and {{Machines}} in {{Wind
  Applications}}}.\hskip 1em plus 0.5em minus 0.4em\relax Denver, CO, USA:
  IEEE, Jul. 2012, pp. 1--8.

\bibitem{moharanaSSRAlleviationSTATCOM2014}
A.~Moharana, R.~Varma, and R.~Seethapathy, ``{{SSR}} alleviation by {{STATCOM}}
  in induction-generator-based wind farm connected to series compensated
  line,'' \emph{IEEE Transactions on Sustainable Energy}, vol.~5, no.~3, pp.
  947--957, 2014.

\bibitem{fanModelingDFIGBasedWind2010}
L.~Fan, R.~Kavasseri, Z.~L. Miao, and C.~Zhu, ``Modeling of {{DFIG-Based Wind
  Farms}} for {{SSR Analysis}},'' \emph{IEEE Transactions on Power Delivery},
  vol.~25, no.~4, pp. 2073--2082, Oct. 2010.

\bibitem{xuSmallSignalStabilityAnalysis2020}
Y.~Xu, M.~Zhang, L.~Fan, and Z.~Miao, ``Small-{{Signal Stability Analysis}} of
  {{Type-4 Wind}} in {{Series-Compensated Networks}},'' \emph{IEEE Transactions
  on Energy Conversion}, vol.~35, no.~1, pp. 529--538, Mar. 2020.

\bibitem{trevisanAnalysisLowFrequency}
A.~S. Trevisan, M.~Fecteau, A.~Mendon{\c c}a, R.~Gagnon, and J.~Mahseredjian,
  ``Analysis of low frequency interactions between {{DFIG}} wind turbines and
  series compensated systems.''

\bibitem{aminSmallSignalStabilityAssessment2017}
M.~Amin and M.~Molinas, ``Small-{{Signal Stability Assessment}} of {{Power
  Electronics Based Power Systems}}: {{A Discussion}} of {{Impedance-}} and
  {{Eigenvalue-Based Methods}},'' \emph{IEEE Transactions on Industry
  Applications}, vol.~53, no.~5, pp. 5014--5030, Sep. 2017.

\bibitem{suriyaarachchiProcedureStudySubSynchronous2013}
D.~H.~R. Suriyaarachchi, U.~D. Annakkage, C.~Karawita, and D.~A. Jacobson, ``A
  {{Procedure}} to {{Study Sub-Synchronous Interactions}} in {{Wind Integrated
  Power Systems}},'' \emph{IEEE Transactions on Power Systems}, vol.~28, no.~1,
  pp. 377--384, Feb. 2013.

\bibitem{raumaResonanceAnalysisWind2012}
K.~Rauma, K.~Md~Hasan, C.~Gavriluta, and C.~Citro, ``Resonance analysis of a
  wind power plant with modal approach,'' in \emph{2012 {{IEEE International
  Symposium}} on {{Industrial Electronics}}}, May 2012, pp. 2042--2047.

\bibitem{liWindWeakGrids2020}
Y.~Li, L.~Fan, and Z.~Miao, ``Wind in {{Weak Grids}}: {{Low-Frequency
  Oscillations}}, {{Subsynchronous Oscillations}}, and {{Torsional
  Interactions}},'' \emph{IEEE Transactions on Power Systems}, vol.~35, no.~1,
  pp. 109--118, Jan. 2020.

\bibitem{bajracharyaUnderstandingTuningTechniques2008}
C.~Bajracharya and M.~Molinas, ``Understanding of tuning techniques of
  converter controllers for {{VSC-HVDC}},'' 2008.

\bibitem{alawasaModelingAnalysisSuppression2013}
K.~M. Alawasa, Y.~A.-R.~I. Mohamed, and W.~Xu, ``Modeling, {{Analysis}}, and
  {{Suppression}} of the {{Impact}} of {{Full-Scale Wind-Power Converters}} on
  {{Subsynchronous Damping}},'' \emph{IEEE Systems Journal}, vol.~7, no.~4, pp.
  700--712, Dec. 2013.

\bibitem{alawasaActiveMitigationSubsynchronous2014}
K.~M. Alawasa, Y.~I. Mohamed, and W.~Xu, ``Active {{Mitigation}} of
  {{Subsynchronous Interactions Between PWM Voltage-Source Converters}} and
  {{Power Networks}},'' \emph{IEEE Transactions on Power Electronics}, vol.~29,
  no.~1, pp. 121--134, Jan. 2014.

\bibitem{hailianxieMitigationSSRPresence2014}
{Hailian Xie} and M.~M. De~Oliveira, ``Mitigation of {{SSR}} in presence of
  wind power and series compensation by {{SVC}},'' in \emph{2014
  {{International Conference}} on {{Power System Technology}}}.\hskip 1em plus
  0.5em minus 0.4em\relax Chengdu: IEEE, Oct. 2014, pp. 2819--2826.

\bibitem{sainzAssessmentSubsynchronousOscillations2019}
L.~Sainz, L.~Monjo, M.~{Cheah-Mane}, and J.~Liang, ``Assessment of
  subsynchronous oscillations in {{AC}} grid-connected {{VSC}} systems with
  type-4 wind turbines,'' \emph{IET Renewable Power Generation}, vol.~13,
  no.~16, pp. 3088--3096, 2019.

\bibitem{fanNyquistStabilityCriterionBasedSSRExplanation2012}
L.~Fan and Z.~Miao, ``Nyquist-{{Stability-Criterion-Based SSR Explanation}} for
  {{Type-3 Wind Generators}},'' \emph{IEEE Transactions on Energy Conversion},
  vol.~27, no.~3, pp. 807--809, Sep. 2012.

\bibitem{chernetOnlineVariationWind2016}
S.~Chernet, M.~Bongiorno, G.~K. Andersen, T.~Lund, and P.~C. Kjaer, ``Online
  variation of wind turbine controller parameter for mitigation of {{SSR}} in
  {{DFIG}} based wind farms,'' in \emph{2016 {{IEEE Energy Conversion
  Congress}} and {{Exposition}} ({{ECCE}})}.\hskip 1em plus 0.5em minus
  0.4em\relax Milwaukee, WI, USA: IEEE, Sep. 2016, pp. 1--8.

\bibitem{aminUnderstandingOriginOscillatory2017}
M.~Amin and M.~Molinas, ``Understanding the {{Origin}} of {{Oscillatory
  Phenomena Observed Between Wind Farms}} and {{HVdc Systems}},'' \emph{IEEE
  Journal of Emerging and Selected Topics in Power Electronics}, vol.~5, no.~1,
  pp. 378--392, Mar. 2017.

\bibitem{liReplicatingRealWorldWind2020a}
Y.~Li, L.~Fan, and Z.~Miao, ``Replicating {{Real-World Wind Farm SSR
  Events}},'' \emph{IEEE Transactions on Power Delivery}, vol.~35, no.~1, pp.
  339--348, Feb. 2020.

\bibitem{nathStudySubSynchronousControl2012}
R.~Nath and C.~{Grande-Moran}, ``Study of {{Sub-Synchronous Control
  Interaction}} due to the interconnection of wind farms to a series
  compensated transmission system,'' in \emph{{{PES T}}\&{{D}} 2012}.\hskip 1em
  plus 0.5em minus 0.4em\relax Orlando, FL, USA: IEEE, May 2012, pp. 1--6.

\bibitem{sahniAdvancedScreeningTechniques2012}
M.~Sahni, D.~Muthumuni, B.~Badrzadeh, A.~Gole, and A.~Kulkarni, ``Advanced
  screening techniques for {{Sub-Synchronous Interaction}} in wind farms,'' in
  \emph{{{PES T}}\&{{D}} 2012}, May 2012, pp. 1--9.

\bibitem{chengReactanceScanCrossoverBased2013}
Y.~Cheng, M.~Sahni, D.~Muthumuni, and B.~Badrzadeh, ``Reactance {{Scan
  Crossover-Based Approach}} for {{Investigating SSCI Concerns}} for
  {{DFIG-Based Wind Turbines}},'' \emph{IEEE Transactions on Power Delivery},
  vol.~28, no.~2, pp. 742--751, Apr. 2013.

\bibitem{guptaFrequencyScanningStudy2013}
S.~Gupta, A.~Moharana, and R.~K. Varma, ``Frequency scanning study of
  sub-synchronous resonance in power systems,'' in \emph{2013 26th {{IEEE
  Canadian Conference}} on {{Electrical}} and {{Computer Engineering}}
  ({{CCECE}})}, May 2013, pp. 1--6.

\bibitem{wenStabilityAnalysisThreephase2014}
B.~Wen, ``Stability {{Analysis}} of {{Three-phase AC Power Systems Based}} on
  {{Measured D-Q Frame Impedances}},'' Ph.D. dissertation, 2014.

\bibitem{renRefinedFrequencyScan2016}
W.~Ren and E.~Larsen, ``A {{Refined Frequency Scan Approach}} to
  {{Sub-Synchronous Control Interaction}} ({{SSCI}}) {{Study}} of {{Wind
  Farms}},'' \emph{IEEE Transactions on Power Systems}, vol.~31, no.~5, pp.
  3904--3912, Sep. 2016.

\bibitem{ryggApparentImpedanceAnalysis2017}
A.~Rygg and M.~Molinas, ``Apparent {{Impedance Analysis}}: {{A Small-Signal
  Method}} for {{Stability Analysis}} of {{Power Electronic-Based Systems}},''
  \emph{IEEE Journal of Emerging and Selected Topics in Power Electronics},
  vol.~5, no.~4, pp. 1474--1486, Dec. 2017.

\bibitem{fanIdentifyingDQDomainAdmittance2021}
L.~Fan, Z.~Miao, P.~Koralewicz, S.~Shah, and V.~Gevorgian, ``Identifying
  {{DQ-Domain Admittance Models}} of a 2.3-{{MVA Commercial Grid-Following
  Inverter}} via {{Frequency-Domain}} and {{Time-Domain Data}},'' \emph{IEEE
  Transactions on Energy Conversion}, vol.~36, no.~3, pp. 2463--2472, Sep.
  2021.

\bibitem{lwinFrequencyScanConsiderations2019}
M.~Lwin, R.~Kazemi, and D.~Howard, ``Frequency {{Scan Considerations}} for
  {{SSCI Analysis}} of {{Wind Power Plants}},'' in \emph{2019 {{IEEE Power}} \&
  {{Energy Society General Meeting}} ({{PESGM}})}, Aug. 2019, pp. 1--5.

\bibitem{liuAnalysisDesignImplementation2020}
Z.~Liu, J.~Liu, and Z.~Liu, ``Analysis, {{Design}}, and {{Implementation}} of
  {{Impulse-Injection-Based Online Grid Impedance Identification With Grid-Tied
  Converters}},'' \emph{IEEE Transactions on Power Electronics}, vol.~35,
  no.~12, pp. 12\,959--12\,976, Dec. 2020.

\bibitem{jacobsComparativeStudyFrequency2023}
K.~Jacobs, Y.~Seyedi, and L.~Meng, ``A comparative study on frequency scanning
  techniques for stability assessment in power systems incorporating wind parks
  {\textbar} {{Elsevier Enhanced Reader}},'' 2023.

\bibitem{mengNewSequenceDomain2023}
L.~Meng, U.~Karaagac, and K.~Jacobs, ``A new sequence domain {{EMT-level}}
  multi-input multi-output frequency scanning method for inverter based
  resources,'' \emph{Electric Power Systems Research}, vol. 220, p. 109312,
  Jul. 2023.

\bibitem{ramakrishnaDQAdmittanceExtraction2023}
R.~H. Ramakrishna, Z.~Miao, L.~Fan, and S.~Shah, ``{{DQ Admittance Extraction}}
  for {{Inverter-Based Resources}}: {{Preprint}},'' \emph{Renewable Energy},
  2023.

\bibitem{shirinzadFrequencyScanBased}
M.~Shirinzad, ``Frequency {{Scan Based Stability Analysis}} of {{Power
  Electronic Systems}},'' Ph.D. dissertation.

\bibitem{bakhshizadehImprovingImpedanceBasedStability2018}
M.~K. Bakhshizadeh, F.~Blaabjerg, J.~Hjerrild, {\L}.~Kocewiak, and C.~L. Bak,
  ``Improving the {{Impedance-Based Stability Criterion}} by {{Using}} the
  {{Vector Fitting Method}},'' \emph{IEEE Transactions on Energy Conversion},
  vol.~33, no.~4, pp. 1739--1747, Dec. 2018.

\bibitem{fanTimeDomainMeasurementBasedDQFrame2021}
L.~Fan and Z.~Miao, ``Time-{{Domain Measurement-Based DQ-Frame Admittance Model
  Identification}} for {{Inverter-Based Resources}},'' \emph{IEEE Transactions
  on Power Systems}, vol.~36, no.~3, pp. 2211--2221, May 2021.

\bibitem{maslennikovDissipatingEnergyFlow2017}
S.~Maslennikov, B.~Wang, and E.~Litvinov, ``Dissipating energy flow method for
  locating the source of sustained oscillations,'' \emph{International Journal
  of Electrical Power \& Energy Systems}, vol.~88, pp. 55--62, Jun. 2017.

\bibitem{chenEnergybasedMethodLocation2013}
L.~Chen, Y.~Min, and W.~Hu, ``An energy-based method for location of power
  system oscillation source,'' \emph{IEEE Transactions on Power Systems},
  vol.~28, no.~2, pp. 828--836, May 2013.

\bibitem{zhuParticipationAnalysisImpedance2022}
Y.~Zhu, Y.~Gu, Y.~Li, and T.~C. Green, ``Participation {{Analysis}} in
  {{Impedance Models}}: {{The Grey-Box Approach}} for {{Power System
  Stability}},'' \emph{IEEE Transactions on Power Systems}, vol.~37, no.~1, pp.
  343--353, Jan. 2022.

\bibitem{aminGrayBoxMethodStability2019}
M.~Amin and M.~Molinas, ``A {{Gray-Box Method}} for {{Stability}} and
  {{Controller Parameter Estimation}} in {{HVDC-Connected Wind Farms Based}} on
  {{Nonparametric Impedance}},'' \emph{IEEE Transactions on Industrial
  Electronics}, vol.~66, no.~3, pp. 1872--1882, Mar. 2019.

\bibitem{shahSequenceImpedanceMeasurement2022}
S.~Shah, P.~Koralewicz, V.~Gevorgian, and R.~Wallen, ``Sequence {{Impedance
  Measurement}} of {{Utility-Scale Wind Turbines}} and {{Inverters}} --
  {{Reference Frame}}, {{Frequency Coupling}}, and {{MIMO}}/{{SISO Forms}},''
  \emph{IEEE Transactions on Energy Conversion}, vol.~37, no.~1, pp. 75--86,
  Mar. 2022.

\bibitem{cespedesImpedanceModelingAnalysis2014}
M.~Cespedes and {Jian Sun}, ``Impedance {{Modeling}} and {{Analysis}} of
  {{Grid-Connected Voltage-Source Converters}},'' \emph{IEEE Transactions on
  Power Electronics}, vol.~29, no.~3, pp. 1254--1261, Mar. 2014.

\bibitem{trevisanAnalyticallyValidatedSSCI2021a}
A.~S. Trevisan, A.~Mendonca, R.~Gagnon, J.~Mahseredjian, and M.~Fecteau,
  ``Analytically {{Validated SSCI Assessment Technique}} for {{Wind Parks}} in
  {{Series Compensated Grids}},'' \emph{IEEE Transactions on Power Systems},
  vol.~36, no.~1, pp. 39--48, Jan. 2021.

\bibitem{ryggModifiedSequenceDomainImpedance2016a}
A.~Rygg, M.~Molinas, C.~Zhang, and X.~Cai, ``A {{Modified Sequence-Domain
  Impedance Definition}} and {{Its Equivalence}} to the dq-{{Domain Impedance
  Definition}} for the {{Stability Analysis}} of {{AC Power Electronic
  Systems}},'' \emph{IEEE Journal of Emerging and Selected Topics in Power
  Electronics}, vol.~4, no.~4, pp. 1383--1396, Dec. 2016.

\bibitem{wangUnifiedImpedanceModel2018}
X.~Wang, L.~Harnefors, and F.~Blaabjerg, ``Unified {{Impedance Model}} of
  {{Grid-Connected Voltage-Source Converters}},'' \emph{IEEE Transactions on
  Power Electronics}, vol.~33, no.~2, pp. 1775--1787, Feb. 2018.

\bibitem{lwinFrequencyScanConsiderations2019b}
M.~Lwin, R.~Kazemi, and D.~Howard, ``Frequency {{Scan Considerations}} for
  {{SSCI Analysis}} of {{Wind Power Plants}},'' in \emph{2019 {{IEEE Power}} \&
  {{Energy Society General Meeting}} ({{PESGM}})}, Aug. 2019, pp. 1--5.

\bibitem{shirinzadFrequencyScanBased2021}
M.~Shirinzad, ``Frequency {{Scan Based Stability Analysis}} of {{Power
  Electronic Systems}},'' Jan. 2021.

\bibitem{francisAlgorithmImplementationSystem2011}
G.~Francis, R.~Burgos, D.~Boroyevich, F.~Wang, and K.~Karimi, ``An algorithm
  and implementation system for measuring impedance in the {{D-Q}} domain,'' in
  \emph{2011 {{IEEE Energy Conversion Congress}} and {{Exposition}}}, Sep.
  2011, pp. 3221--3228.

\bibitem{aleniusAmplitudeDesignPerturbation2020}
H.~Alenius, R.~Luhtala, and T.~Roinila, ``Amplitude {{Design}} of
  {{Perturbation Signal}} in {{Frequency-Domain Analysis}} of {{Grid-Connected
  Systems}},'' \emph{IFAC-PapersOnLine}, vol.~53, no.~2, pp. 13\,161--13\,166,
  Jan. 2020.

\bibitem{martinezPerturbationAnalysisPower2004}
I.~Mart{\'{\i}}nez, A.~R. Messina, and E.~Barocio, ``Perturbation analysis of
  power systems: Effects of second- and third-order nonlinear terms on system
  dynamic behavior,'' \emph{Electric Power Systems Research}, vol.~71, no.~2,
  pp. 159--167, Oct. 2004.

\bibitem{badrzadehSubsynchronousInteractionWind2012a}
B.~Badrzadeh and Y.~Zhou, ``Sub-synchronous {{Interaction}} in {{Wind Power
  Plants- Part I}}: {{Study Tools}} and {{Techniques}},'' in \emph{2012 {{IEEE
  Power}} and {{Energy Society General Meeting}}}.\hskip 1em plus 0.5em minus
  0.4em\relax San Diego, CA: IEEE, Jul. 2012, pp. 1--9.

\bibitem{zhangResearchSynchronousControl2020}
Z.~Zhang, P.~Wang, F.~Gao, Z.~Liu, J.~Zhang, and P.~Jiang, ``Research on
  {{Synchronous Control Method}} for {{Suppressing Nonlinear Impulse
  Perturbation}} of {{Photovoltaic Grid-Connected Inverter}},'' \emph{IEEE
  Access}, vol.~8, pp. 22\,303--22\,313, 2020.

\bibitem{foyenImpedanceScanningChirps2022}
S.~Foyen, C.~Zhang, M.~Molinas, O.~Fosso, and T.~Isobe, ``Impedance scanning
  with chirps for single-phase converters,'' in \emph{2022 {{International
  Power Electronics Conference}} ({{IPEC-Himeji}} 2022- {{ECCE Asia}})}.\hskip
  1em plus 0.5em minus 0.4em\relax Himeji, Japan: IEEE, May 2022, pp. 213--219.

\bibitem{slepskiOptimizationImpedanceMeasurements2009}
P.~Slepski and K.~Darowicki, ``Optimization of impedance measurements using
  `chirp' type perturbation signal,'' \emph{Measurement}, vol.~42, no.~8, pp.
  1220--1225, Oct. 2009.

\bibitem{nercIntegratingInverterBasedResources2017}
NERC, ``Integrating {{Inverter-Based Resources}} into {{Low Short Circuit
  Strength Systems Reliability Guideline}},'' Dec. 2017.

\bibitem{mateu-barriendosOscillatoryFrequencyCharacterization2023}
E.~{Mateu-Barriendos}, M.~{Cheah-Mane}, E.~{Prieto-Araujo}, H.~Mehrjerdi, and
  O.~{Gomis-Bellmunt}, ``Oscillatory frequency characterization based on
  impedance analysis,'' \emph{International Journal of Electrical Power \&
  Energy Systems}, vol. 152, p. 109208, Oct. 2023.

\bibitem{yoonHarmonicStabilityAssessment2016}
C.~Yoon, H.~Bai, R.~N. Beres, X.~Wang, C.~L. Bak, and F.~Blaabjerg, ``Harmonic
  {{Stability Assessment}} for {{Multiparalleled}}, {{Grid-Connected
  Inverters}},'' \emph{IEEE Transactions on Sustainable Energy}, vol.~7, no.~4,
  pp. 1388--1397, Oct. 2016.

\bibitem{wangModelingAnalysisHarmonic2014}
X.~Wang, F.~Blaabjerg, and W.~Wu, ``Modeling and {{Analysis}} of {{Harmonic
  Stability}} in an {{AC Power-Electronics-Based Power System}},'' \emph{IEEE
  Transactions on Power Electronics}, vol.~29, no.~12, pp. 6421--6432, Dec.
  2014.

\bibitem{zhangImpedanceBasedLocalStability2015}
X.~Zhang, X.~Ruan, and C.~K. Tse, ``Impedance-{{Based Local Stability
  Criterion}} for {{DC Distributed Power Systems}},'' \emph{IEEE Transactions
  on Circuits and Systems I: Regular Papers}, vol.~62, no.~3, pp. 916--925,
  Mar. 2015.

\bibitem{jiansunSmallSignalMethodsAC2009}
{Jian Sun}, ``Small-{{Signal Methods}} for {{AC Distributed Power
  Systems}}--{{A Review}},'' \emph{IEEE Transactions on Power Electronics},
  vol.~24, no.~11, pp. 2545--2554, Nov. 2009.

\bibitem{shahReversedImpedancebasedStability2023}
S.~Shah, W.~Yan, P.~Koralewicz, E.~Mendiola, and V.~Gevorgian, ``A reversed
  impedance-based stability criterion for {{IBR}} grids,'' \emph{IET Conference
  Proceedings}, vol. 2022, no.~23, pp. 157--164, Jan. 2023.

\bibitem{wengkhuenhoDirectNyquistArray2000}
{Weng Khuen Ho}, {Tong Heng Lee}, {Wen Xu}, J.~Zhou, and {Ee Beng Tay}, ``The
  direct {{Nyquist}} array design of {{PID}} controllers,'' \emph{IEEE
  Transactions on Industrial Electronics}, vol.~47, no.~1, pp. 175--185, Feb.
  2000.

\bibitem{turnerCaseStudyApplication2013}
R.~Turner, S.~Walton, and R.~Duke, ``A {{Case Study}} on the {{Application}} of
  the {{Nyquist Stability Criterion}} as {{Applied}} to {{Interconnected
  Loads}} and {{Sources}} on {{Grids}},'' \emph{IEEE Transactions on Industrial
  Electronics}, vol.~60, no.~7, pp. 2740--2749, Jul. 2013.

\bibitem{ryggModifiedSequenceDomainImpedance2016}
A.~Rygg, M.~Molinas, C.~Zhang, and X.~Cai, ``A {{Modified Sequence-Domain
  Impedance Definition}} and {{Its Equivalence}} to the dq-{{Domain Impedance
  Definition}} for the {{Stability Analysis}} of {{AC Power Electronic
  Systems}},'' \emph{IEEE Journal of Emerging and Selected Topics in Power
  Electronics}, vol.~4, no.~4, pp. 1383--1396, Dec. 2016.

\bibitem{aminNyquistStabilityCriterion2019}
M.~Amin, C.~Zhang, A.~Rygg, M.~Molinas, E.~Unamuno, and M.~Belkhayat, ``Nyquist
  {{Stability Criterion}} and its {{Application}} to {{Power Electronics
  Systems}},'' in \emph{Wiley {{Encyclopedia}} of {{Electrical}} and
  {{Electronics Engineering}}}, 1st~ed., J.~G. Webster, Ed.\hskip 1em plus
  0.5em minus 0.4em\relax Wiley, May 2019, pp. 1--22.

\bibitem{shahReversedImpedancebasedStability2023a}
S.~Shah, W.~Yan, P.~Koralewicz, E.~Mendiola, and V.~Gevorgian, ``A reversed
  impedance-based stability criterion for {{IBR}} grids,'' \emph{IET Conference
  Proceedings}, vol. 2022, no.~23, pp. 157--164, Jan. 2023.

\bibitem{liuSubsynchronousInteractionDirectDrive2017c}
H.~Liu, X.~Xie, J.~He, T.~Xu, Z.~Yu, C.~Wang, and C.~Zhang, ``Subsynchronous
  {{Interaction Between Direct-Drive PMSG Based Wind Farms}} and {{Weak AC
  Networks}},'' \emph{IEEE Transactions on Power Systems}, vol.~32, no.~6, pp.
  4708--4720, Nov. 2017.

\bibitem{qiaoSmallSignalStabilityAnalysis2024}
L.~Qiao, Y.~Xue, L.~Kong, F.~Wang, and {Nupur}, ``Small-{{Signal Stability
  Analysis}} for {{Large-Scale Power Electronics- Based Power Systems}},''
  \emph{IEEE Open Access Journal of Power and Energy}, vol.~11, pp. 280--292,
  2024.

\bibitem{liStabilityAnalysisLocation2021}
Y.~Li, Z.~Shuai, X.~Liu, Y.~Chen, Z.~Li, Y.~Hong, and Z.~J. Shen, ``Stability
  {{Analysis}} and {{Location Optimization Method}} for {{Multiconverter Power
  Systems Based}} on {{Nodal Admittance Matrix}},'' \emph{IEEE Journal of
  Emerging and Selected Topics in Power Electronics}, vol.~9, no.~1, pp.
  529--538, Feb. 2021.

\end{thebibliography}

\end{document}